%% file: main.tex
\documentclass[twoside]{article}

\usepackage[accepted]{aistats2024}
\usepackage[round]{natbib}

\usepackage{hyperref}
\usepackage[capitalize,noabbrev]{cleveref}
\usepackage{graphicx}
\newcommand{\twocolumnstyle}{}
\usepackage{booktabs}   
\makeatletter

\usepackage{xcolor}
\newcommand{\newtext}[1]{#1}
\newcommand{\cameraready}[1]{#1}
\usepackage{amsthm}
\newtheorem{theorem}{Theorem} 
\newtheorem{proposition}{Proposition}
\usepackage{mathrsfs}
\usepackage{amsfonts}
\usepackage{graphicx}
\usepackage{subcaption}
\usepackage{tikz}
\theoremstyle{definition}
\usepackage{url}

\begin{document}

%

%

\twocolumn[

\aistatstitle{Integrating Uncertainty Awareness into Conformalized Quantile Regression}

\aistatsauthor{ Raphael Rossellini \And Rina Foygel Barber \And  Rebecca Willett }

\aistatsaddress{ University of Chicago \And  University of Chicago \And University of Chicago } ]


\begin{abstract}
\input{Chapters/Abstract}
\end{abstract}

\input{Chapters/Introduction}
\input{Chapters/Existing_methods}
\input{Chapters/Uncertainty_aware_CQR}

\input{Chapters/Simulation_case_study}

\input{Chapters/Performance_on_real_world_data_sets}
\input{Chapters/Discussion_and_limitations}
\input{Chapters/Acknowledgements}

\bibliography{main.bib}





\section*{Checklist}



 \begin{enumerate}

 \item For all models and algorithms presented, check if you include:
 \begin{enumerate}
   \item A clear description of the mathematical setting, assumptions, algorithm, and/or model. 

    Yes, see \cref{sec:existing-methods} and \cref{sec:UACQR}.
   
   \item An analysis of the properties and complexity (time, space, sample size) of any algorithm. 

   Yes, see \cref{apdx:computation}.
   
   \item (Optional) Anonymized source code, with specification of all dependencies, including external libraries. 

   Yes
 \end{enumerate}

 \item For any theoretical claim, check if you include:
 \begin{enumerate}
   \item Statements of the full set of assumptions of all theoretical results. 

   Yes, see \cref{sec:theory}
   
   \item Complete proofs of all theoretical results.

   Yes, see \cref{sec:theory} and \cref{apdx:theory}
   
   \item Clear explanations of any assumptions. 

   Yes, see sections above.
 \end{enumerate}

 \item For all figures and tables that present empirical results, check if you include:
 \begin{enumerate}
   \item The code, data, and instructions needed to reproduce the main experimental results (either in the supplemental material or as a URL). 

    Yes, code for reproducibility is included in supplementary material.
   
   \item All the training details (e.g., data splits, hyperparameters, how they were chosen). 

    Yes, these details are included in \cref{apdx:computation}, \cref{apdx:methods}, and in provided source code
   
         \item A clear definition of the specific measure or statistics and error bars (e.g., with respect to the random seed after running experiments multiple times).

        Yes, see \cref{sec:real-world}.
         
         \item A description of the computing infrastructure used. (e.g., type of GPUs, internal cluster, or cloud provider).

         Yes, see \cref{apdx:methods}.
 \end{enumerate}

 \item If you are using existing assets (e.g., code, data, models) or curating/releasing new assets, check if you include:
 \begin{enumerate}
   \item Citations of the creator If your work uses existing assets. 

    Yes, see end of \cref{apdx:methods}.
   \item The license information of the assets, if applicable.

    Not applicable.
   
   \item New assets either in the supplemental material or as a URL, if applicable. 

   Not applicable.
   \item Information about consent from data providers/curators.

   Not applicable.
   \item Discussion of sensible content if applicable, e.g., personally identifiable information or offensive content.

   Not applicable.
 \end{enumerate}

 \item If you used crowdsourcing or conducted research with human subjects, check if you include:
 \begin{enumerate}
   \item The full text of instructions given to participants and screenshots. 

   Not applicable.
   \item Descriptions of potential participant risks, with links to Institutional Review Board (IRB) approvals if applicable. 

   Not applicable.
   \item The estimated hourly wage paid to participants and the total amount spent on participant compensation.

   Not applicable.
 \end{enumerate}

 \end{enumerate}

\clearpage
\newpage
\onecolumn
\appendix

\input{Chapters/Fast_heuristic_alternatives}

\input{Chapters/Methodological_details}
\input{Chapters/Randomized_cutoffs}
\input{Chapters/Lipschitz_property}

\input{Chapters/Additional_experiments}

\end{document}

%% file: Chapters/Abstract.tex
  Conformalized Quantile Regression (CQR) is a recently proposed method for constructing prediction intervals for  a response $Y$ given covariates $X$, without making distributional assumptions. However, existing constructions of CQR can be ineffective for problems where the quantile regressors perform better in certain parts of the feature space than others. The reason is that the prediction intervals of CQR do not distinguish between two forms of uncertainty: first, the variability of the conditional distribution of $Y$ given $X$ (i.e., aleatoric uncertainty), and second, our uncertainty in estimating this conditional distribution (i.e., epistemic uncertainty). This can lead to intervals that are overly narrow in regions where epistemic uncertainty is high. To address this, we propose a new variant of the CQR methodology, Uncertainty-Aware CQR (UACQR), that explicitly separates these two sources of uncertainty to adjust quantile regressors differentially across the feature space. Compared to CQR, our methods enjoy the same distribution-free theoretical coverage guarantees, while demonstrating in our experiments stronger conditional coverage properties in simulated settings and real-world data sets alike.

%% file: Chapters/Introduction.tex
\section{INTRODUCTION}

While machine learning approaches in recent years have frequently succeeded in producing models that produce predictions with low error on average, there are many instances in which these models can provide a false sense of precision. Prediction intervals are one of the most natural ways to quantify uncertainty, providing bounds on the true response that hold with some desired probability, frequently 90\%. To this end, conformal prediction has emerged as an attractive paradigm for creating statistically-valid prediction intervals for black-box machine learning models with no distributional assumptions. Specifically, suppose we have observed $n$ samples $(X_i, Y_i) \in \mathbb{R}^p \times \mathbb{R}$, and we desire a prediction interval for $Y_{n+1}$ based on $X_{n+1}$. If the $n+1$ samples are exchangeable (e.g., if the pairs $(X_i,Y_i)$ are drawn i.i.d.\ from an arbitrary shared distribution), conformal prediction allows one to construct an interval $\hat{C}_n(X_{n+1})$ such that 
\begin{equation}
  \mathbb{P}\left\{Y_{n+1} \in \hat{C}_n(X_{n+1}) \right\} \geq 1-\alpha \label{eqn:marginal_coverage}  
\end{equation}
A remaining challenge is the ability to provide prediction intervals that hold conditionally on the test point’s covariates, 
\[
\mathbb{P}\left\{Y_{n+1} \in \hat{C}_n(X_{n+1}) \mid X_{n+1}=x \right\} \geq 1-\alpha,
\]
which is not ensured by conformal prediction under the exchangeability assumption alone.
Unfortunately, it is impossible for any distribution-free approach to guarantee conditional coverage in a non-trivial way when the covariates follow a continuous distribution \citep{vovk_book,lei2014distribution,barber2021limits}.

In spite of this impossibility result, there have been numerous methodological advances that allow conformal prediction to provide better conditional coverage properties empirically. A particularly notable advancement has been Conformalized Quantile Regression (CQR) \citep{Romano2019ConformalizedQR}, which uses quantile regressors as inputs into the conformal prediction paradigm to create prediction intervals that adapt to the covariates. Part of the appeal is that the true conditional quantile functions are a sufficient ingredient to achieving conditional coverage. If we let $q_{Y \mid X}(x; \alpha_{\textnormal{lo}}),\; q_{Y \mid X}(x; \alpha_{\textnormal{hi}})$ be the \textit{true} $\alpha_{\textnormal{lo}}, \alpha_{\textnormal{hi}}$ conditional quantiles for $Y\mid X=x$, where $\alpha_{\textnormal{lo}}=\frac{\alpha}{2}$ and $\alpha_{\textnormal{hi}} = 1-\frac{\alpha}{2}$, then by construction
\[
\mathbb{P} \left\{Y \in [q_{Y \mid X}(X; \alpha_{\textnormal{lo}}), q_{Y \mid X}(X; \alpha_{\textnormal{hi}})] \mid X=x \right\} \geq 1-\alpha.
\]
Thus, if one can find accurate quantile regressors via machine learning, CQR holds potential to create prediction intervals that approximately have conditional coverage while maintaining the marginal validity of previous conformal prediction methods.
Specifically, for any $a\in(0,1)$, let $\hat q_{Y \mid X}(x,a)$ denote an estimate of the conditional  $a$-quantile of $Y \mid X=x$ computed using training data. CQR generates prediction intervals of the form
\[
\hat{C}_n(X_{n+1}) =[\hat{q}_{Y \mid X}(X_{n+1}; \alpha_{\textnormal{lo}})-t, \hat{q}_{Y \mid X}(X_{n+1}; \alpha_{\textnormal{hi}})+t]
\]
where calibration data is selected to choose $t$.

In practice, however, our estimate $\hat{q}_{Y \mid X}(x,a)$ may be less accurate in particularly challenging regions of the feature space. CQR struggles in this setting since it simply inflates the initial estimates by a constant, non-adaptive, additive adjustment $t \in \mathbb{R}$.

Others have implemented adaptive variants. For example, the CQR-r variant from \citet{sesia2020comparison} multiplies $t$ by the interval width $w(x) = \hat{q}_{Y \mid X}(x; \alpha_{\textnormal{hi}}) - \hat{q}_{Y \mid X}(x; \alpha_{\textnormal{lo}})$, which yields: 
\ifdefined\twocolumnstyle
\begin{align*}
    \hat{C}_n(X_{n+1}) =[&\hat{q}_{Y \mid X}(X_{n+1}; \alpha_{\textnormal{lo}})-t\cdot w(X_{n+1}),\\ 
    &\hat{q}_{Y \mid X}(X_{n+1}; \alpha_{\textnormal{hi}})+t\cdot w(X_{n+1})]
\end{align*}
\else
  \[
\hat{C}_n(X_{n+1}) =[\hat{q}_{Y \mid X}(X_{n+1}; \alpha_{\textnormal{lo}})-t\cdot w(X_{n+1}), \hat{q}_{Y \mid X}(X_{n+1}; \alpha_{\textnormal{hi}})+t\cdot w(X_{n+1})]
\]
\fi

The issue here is that this is scaling $t$ by the wrong type of uncertainty. 
$w(x)$ is a proxy for \textit{aleatoric uncertainty}---the irreducible uncertainty inherent to a data generating process, i.e., the amount of variability in the conditional distribution of $Y \mid X$.
However, the gap between the upper and lower quantile regressors already accounts for aleatoric uncertainty. The additive adjustments should ideally reflect the \textit{epistemic uncertainty}---the scale of the estimation error of the quantile regressors at each $X$\newtext{:
\begin{equation*}
    |\hat{q}_{Y \mid X}(x,a) - q_{Y \mid X}(x,a)|, \; a \in \{\alpha_{\textnormal{lo}}, \alpha_{\textnormal{hi}} \}
\end{equation*}}
The aleatoric uncertainty may be associated with epistemic uncertainty in many contexts, but there is no guarantee that there must be a correspondence between the two. In medical studies, for instance, a poorly-represented demographic could have a narrower range possible of outcomes than a well-represented one, meaning that this subpopulation has high epistemic uncertainty yet low aleatoric uncertainty. Our work centers on the idea that, for this type of setting, the quantile-based prediction interval should be inflated in a way that scales with how well sampled each demographic is (epistemic uncertainty), rather than with the estimated range of outcomes for each demographic (aleatoric uncertainty).

\subsection{Our contribution}
We propose a new family of methods for conformalizing quantile regression that incorporate epistemic uncertainty.  We call them \textit{UACQR-S (Uncertainty-Aware CQR via Scaling)} and \textit{UACQR-P (Uncertainty-Aware CQR via Percentiles)}
\footnote{We provide code implementing UACQR at \href{https://github.com/rrross/UACQR}{https://github.com/rrross/UACQR}}
UACQR-S creates prediction intervals of a similar form to CQR-r, but its scaling function directly estimates the epistemic, rather than aleatoric, uncertainty in $Y \mid X$ at each value of $X$. In contrast, UACQR-P is constructed based on an ensemble of quantile regression estimates (e.g., obtained via bootstrapping the training sample), and outputs a prediction interval that is defined as a  calibrated percentile of the $B$ different estimates at each point.
Our methods offer the same distribution-free guarantee of validity as existing CQR constructions, while offering empirical improvements in terms of conditional coverage.

To minimize computational concerns, we provide recipes for implementing UACQR-S and UACQR-P that have the same computational cost of CQR when using any one of a broad range of machine learning paradigms, namely bagging (which includes Random Forests)  and epoch-based optimization (e.g., neural networks). These two classes cover a large proportion of successful modern machine learning architectures, so our proposed methods hold potential for widespread applications. 

%% file: Chapters/Existing_methods.tex
\section{EXISTING METHODS\label{sec:existing-methods}}

We  review split conformal prediction \citep{vovk_book} in more depth, and  define a generic meta-algorithm that captures all existing variants of CQR. 

Suppose we are given a data set $(X_1,Y_1),\dots,(X_n,Y_n)$ and a new test point $X_{n+1}$ for which we would like to predict the response, $Y_{n+1}$.  We limit ourselves to assuming only that $(X_1,Y_1),\dots,(X_{n+1},Y_{n+1})$ are exchangeable. In its most basic version, split conformal prediction operates as follows. We split the $n$ data points into a training set and calibration set of sizes $n_0+n_1=n$. After fitting a predictive model $\hat\mu$ to the training set $\{(X_i,Y_i):i=1,\dots,n_0\}$, we then define
\begin{equation}\label{eqn:split_conformal_simple}\widehat{C}_n(X_{n+1}) = \hat\mu(X_{n+1})\pm \hat t,\end{equation}
where $\hat t$ is the $\lceil(1-\alpha)(n_1+1)\rceil$-th smallest element of the calibration set residuals $\{|Y_i - \hat\mu(X_i)|:i=n_0+1,\dots,n\}$. 

Split conformal prediction is much more general than this simple construction, however, and in particular it can be used to produce prediction intervals of any ``shape'' rather than restricting ourselves only to constant-width type intervals of the form $\hat\mu(x)\pm \textnormal{(constant)}$. \citet{gupta2022nested} describe a formulation based on nested sets: suppose that, given the training data $\{(X_i,Y_i):i=1,\dots,n_0\}$, we construct a family of nested prediction bands $\hat{C}_{n_0,t}$ (where $\hat{C}_{n_0,t}(x)\subseteq\mathbb{R}$ is a prediction set or prediction interval at each $x$), indexed over some set $t\in\mathcal{T}\subseteq\mathbb{R}$. This family is required to be nested, with $\hat{C}_{n_0,t}(x)\subseteq\hat{C}_{n_0,t'}(x)$ for $t<t'$, and also that $\inf_{t\in\mathcal{T}}\hat{C}_{n_0,t}(x)=\emptyset$ and $\sup_{t\in\mathcal{T}}\hat{C}_{n_0,t}(x)=\mathbb{R}$.
Then, using the calibration set, define
\begin{equation}
\hat{t} = \inf\left\{ t: \sum_{i=n_0+1}^n \mathbf{1}_{Y_i \in \hat{C}_{n_0,t}(X_i)}\geq (1-\alpha)(n_1+1)\right\},
\label{eq:nested_that}
\end{equation}
and return the prediction set $\hat{C}_n(X_{n+1}) : = \hat{C}_{n_0,\hat{t}}(X_{n+1})$. To summarize, the training points ($i=1,\dots,n_0$) define a nested family of prediction bands $\{\hat{C}_{n_0,t}\}$, and then the calibration set ($i=n_0+1,\dots,n_1$) is used to choose a particular $\hat{t}$ that selects one prediction band from this family to then provide at least $1-\alpha$ coverage of the final prediction interval.
\citet{vovk_book} and \citet{gupta2022nested} show any construction of this form offers the distribution-free prediction guarantee~\eqref{eqn:marginal_coverage}.
To see how this general formulation applies to the simple construction~\eqref{eqn:split_conformal_simple} given above, we can simply take $\hat{C}_t(x) = \hat\mu(x) \pm t$ for $t\in\mathcal{T}=\mathbb R$ to recover the prediction set given in~\eqref{eqn:split_conformal_simple}. 

To relate this construction to the problem of quantile regression, we  change the notation to provide prediction sets that are specifically designed to be intervals. 
Let $\hat{q}_{\textnormal{lo}}(X, t)$ and $\hat{q}_{\textnormal{hi}}(X, t)$ be quantile regressors for $Y \mid X$, fitted using the training data $\{(X_i,Y_i):i=1,\dots,n_0\}$, where $t\mapsto \hat q_{\textnormal{lo}}(x,t)$ is decreasing, while $t\mapsto \hat q_{\textnormal{hi}}(x,t)$ is increasing.  We  consider the nested family of sets given by $\hat{C}_{n_0,t}(x) = [\hat{q}_{\textnormal{lo}}(x,t),\hat{q}_{\textnormal{hi}}(x,t)]$, and consequently, split conformal prediction returns the interval
\begin{equation}\label{eqn:nested_q_C_hat}\hat{C}_n(X_{n+1}) = \hat{C}_{n_0,\hat{t}}(X_{n+1}) = [ \hat{q}_{\textnormal{lo}}(X_{n+1},\hat{t}),\hat{q}_{\textnormal{hi}}(X_{n+1},\hat{t})]\end{equation}
where, for this specific construction, the definition~\eqref{eq:nested_that} of $\hat{t}$ simplifies to
\ifdefined\twocolumnstyle
\begin{equation}\label{eqn:nested_q_t_hat}\inf\left\{ t: \sum_{i=n_0+1}^n \mathbf{1}_{\hat{q}_{\textnormal{lo}}(X_i,t)\leq Y_i \leq \hat{q}_{\textnormal{hi}}(X_i,t)}\geq (1-\alpha)(n_1+1)\right\}\end{equation}
\else
\begin{equation}\label{eqn:nested_q_t_hat}\hat{t} = \inf\left\{ t: \sum_{i=n_0+1}^n \mathbf{1}_{\hat{q}_{\textnormal{lo}}(X_i,t)\leq Y_i \leq \hat{q}_{\textnormal{hi}}(X_i,t)}\geq (1-\alpha)(n_1+1)\right\}.\end{equation}
\fi
(To satisfy the earlier conditions, we also assume that $ \inf_{t\in\mathcal{T}}\hat{q}_{\textnormal{lo}}(x,t)=-\infty$ and $ \sup_{t\in\mathcal{T}}\hat{q}_{\textnormal{lo}}(x,t)=+\infty$, and $ \inf_{t\in\mathcal{T}}\hat{q}_{\textnormal{hi}}(x,t)=-\infty$ and $ \sup_{t\in\mathcal{T}}\hat{q}_{\textnormal{hi}}(x,t)=+\infty$.)

Next we  review CQR and several related existing methods. Our presentation reformulates each method using the general notation we have just defined.

\paragraph{Conformalized Quantile Regression (CQR)}
For any $a\in(0,1)$, let $\hat{q}_{Y \mid X}(x,a)$ denote an estimate (fitted on the training data set) of the conditional $a$-quantile of $Y$ given $X=x$. \citet{Romano2019ConformalizedQR} proposed fitting quantile regressors for preset upper and lower quantiles, $\alpha_{\textnormal{lo}}$ and $\alpha_{\textnormal{hi}}$ (e.g., $\alpha_{\textnormal{lo}}=\alpha/2$ and $\alpha_{\textnormal{hi}}=1-\alpha/2$), and then using conformal prediction to select a constant additive adjustment to expand or contract these intervals to ensure marginal coverage. This produces a prediction interval of the form
\[\hat{C}_n(X_{n+1}) = [\hat{q}_{Y \mid X}(X_{n+1},\alpha_{\textnormal{lo}}) - \hat{t}, \hat{q}_{Y \mid X}(X_{n+1},\alpha_{\textnormal{hi}}) + \hat{t}].\]
By choosing
\[\hat{q}_{\textnormal{lo}}(x, t) = \hat{q}_{Y \mid X}(x, \alpha_{\textnormal{lo}}) - t,\quad
    \hat{q}_{\textnormal{hi}}(x, t) = \hat{q}_{Y \mid X}(x, \alpha_{\textnormal{hi}}) + t,
\]
indexed by $t\in\mathbb{R}$,
we can see that the CQR  interval is equal to the general construction~\eqref{eqn:nested_q_C_hat} above.

\paragraph{CQR-r and CQR-m} \citet{sesia2020comparison} propose a variant of CQR, which they call CQR-r. The prediction interval is given by
\ifdefined\twocolumnstyle
\begin{align*}
    \hat{C}_n(X_{n+1}) =[&\hat{q}_{Y \mid X}(X_{n+1}; \alpha_{\textnormal{lo}})-t\cdot w(X_{n+1}),\\ 
    &\hat{q}_{Y \mid X}(X_{n+1}; \alpha_{\textnormal{hi}})+t\cdot w(X_{n+1})]\\
    w(x):=&\hat{q}_{Y \mid X}(x; \alpha_{\textnormal{hi}}) - \hat{q}_{Y \mid X}(x; \alpha_{\textnormal{lo}})
\end{align*}
\else
\begin{multline*}\hat{C}_n(X_{n+1}) = [\hat{q}_{Y \mid X}(X_{n+1},\alpha_{\textnormal{lo}}) - \hat{t}\cdot (\hat{q}_{Y \mid X}(X_{n+1}, \alpha_{\textnormal{hi}}) - \hat{q}_{Y \mid X}(X_{n+1}, \alpha_{\textnormal{lo}})),\\ \hat{q}_{Y \mid X}(X_{n+1},\alpha_{\textnormal{hi}}) + \hat{t}\cdot (\hat{q}_{Y \mid X}(X_{n+1}, \alpha_{\textnormal{hi}}) - \hat{q}_{Y \mid X}(X_{n+1}, \alpha_{\textnormal{lo}}))],\end{multline*}
\fi
which we can interpret as scaling the inflation of the interval by the \emph{aleatoric} uncertainty in $Y \mid X$.
This method corresponds to the general construction \eqref{eqn:nested_q_C_hat} above obtained by using
\ifdefined\twocolumnstyle
\begin{align*}
    &\hat{q}_{\textnormal{lo}}(x, t) = \hat{q}_{Y \mid X}(x, \alpha_{\textnormal{lo}}) - t\cdot w(x),\\
    &\hat{q}_{\textnormal{hi}}(x, t) = \hat{q}_{Y \mid X}(x, \alpha_{\textnormal{hi}}) + t\cdot w(x),
\end{align*}
\else
\begin{align*}
    &\hat{q}_{\textnormal{lo}}(x, t) = \hat{q}_{Y \mid X}(x, \alpha_{\textnormal{lo}}) - t\cdot (\hat{q}_{Y \mid X}(x, \alpha_{\textnormal{hi}}) - \hat{q}_{Y \mid X}(x, \alpha_{\textnormal{lo}})),\\
    &\hat{q}_{\textnormal{hi}}(x, t) = \hat{q}_{Y \mid X}(x, \alpha_{\textnormal{hi}}) + t\cdot (\hat{q}_{Y \mid X}(x, \alpha_{\textnormal{hi}}) - \hat{q}_{Y \mid X}(x, \alpha_{\textnormal{lo}})),
\end{align*}
\fi
where $t\in\mathbb{R}$.
\citet{sesia2020comparison}'s CQR-r construction is a variant of the method proposed by \citet{kivaranovic2020adaptive} (called ``CQR-m'' by \citet{sesia2020comparison}), where the scaling factor instead compares the (estimated) upper and lower quantiles to the (estimated) median:
\ifdefined\twocolumnstyle
\begin{align*}
    &\hat{q}_{\textnormal{lo}}(x, t) = \hat{q}_{Y \mid X}(x, \alpha_{\textnormal{lo}}) - t\cdot w_{\textnormal{lo}}(x),\\
   &\hat{q}_{\textnormal{hi}}(x, t) = \hat{q}_{Y \mid X}(x, \alpha_{\textnormal{hi}}) + t\cdot w_{\textnormal{hi}}(x),\\
   &w_{\textnormal{lo}}(x) :=\hat{q}_{Y \mid X}(x, 0.5) - \hat{q}_{Y \mid X}(x, \alpha_{\textnormal{lo}}),\\
   &w_{\textnormal{hi}}(x) :=\hat{q}_{Y \mid X}(x, \alpha_{\textnormal{hi}}) - \hat{q}_{Y \mid X}(x, 0.5),
\end{align*}
\else
\begin{align*}
    &\hat{q}_{\textnormal{lo}}(x, t) = \hat{q}_{Y \mid X}(x, \alpha_{\textnormal{lo}}) - t\cdot (\hat{q}_{Y \mid X}(x, 0.5) - \hat{q}_{Y \mid X}(x, \alpha_{\textnormal{lo}})),\\
   &\hat{q}_{\textnormal{hi}}(x, t) = \hat{q}_{Y \mid X}(x, \alpha_{\textnormal{hi}}) + t\cdot (\hat{q}_{Y \mid X}(x, \alpha_{\textnormal{hi}}) - \hat{q}_{Y \mid X}(x, 0.5)),
\end{align*}
\fi
again with $t\in\mathbb{R}$, leading to the prediction interval
\ifdefined\twocolumnstyle
\begin{align*}\hat{C}_n(X_{n+1}) = [&\hat{q}_{Y \mid X}(X_{n+1},\alpha_{\textnormal{lo}}) - \hat{t}\cdot w_{\textnormal{lo}}(X_{n+1}),\\ &\hat{q}_{Y \mid X}(X_{n+1},\alpha_{\textnormal{hi}}) + \hat{t}\cdot w_{\textnormal{hi}}(X_{n+1})].\end{align*}
\else
\begin{multline*}\hat{C}_n(X_{n+1}) = [\hat{q}_{Y \mid X}(X_{n+1},\alpha_{\textnormal{lo}}) - \hat{t}\cdot (\hat{q}_{Y \mid X}(X_{n+1}, 0.5) - \hat{q}_{Y \mid X}(X_{n+1}, \alpha_{\textnormal{lo}})),\\ \hat{q}_{Y \mid X}(X_{n+1},\alpha_{\textnormal{hi}}) + \hat{t}\cdot (\hat{q}_{Y \mid X}(X_{n+1}, \alpha_{\textnormal{hi}}) - \hat{q}_{Y \mid X}(X_{n+1}, 0.5))].\end{multline*}
\fi

\paragraph{Distributional Conformal Prediction (DCP)} An alternative style of approach is proposed by \citet{chernozhukov2021distributional}, which uses conformal prediction to select the target upper and lower quantiles for quantile regression to achieve the desired coverage. After fitting the conditional $t$-quantile $\hat{q}_{Y \mid X}(x,t)$ for every $t\in\mathcal{T}\subseteq[0,1]$, the prediction set is given by
\[\hat{C}_n(X_{n+1}) = [\hat{q}_{Y \mid X}(X_{n+1},\hat{t}), \hat{q}_{Y \mid X}(X_{n+1},1-\hat{t})], \]
which corresponds to the general construction \eqref{eqn:nested_q_C_hat} above by using the following, with $t \in \mathcal{T}\subseteq[0,1]$:
\[
   \hat{q}_{\textnormal{lo}}(x, t) = \hat{q}_{Y \mid X}(x,t),\quad
   \hat{q}_{\textnormal{hi}}(x, t) = \hat{q}_{Y \mid X}(x,1-t).
\]

\subsection{Limitations of existing methods}
To understand the motivation behind our new methods, we briefly consider the limitations of this existing range of constructions. Consider the ``baseline'' prediction interval $\hat{C}_{\textnormal{base}}(x) = [\hat{q}_{Y \mid X}(x,\alpha/2),\hat{q}_{Y \mid X}(x,1-\alpha/2)]$, given by running quantile regression on the training set. In many settings, this interval may more closely resemble the oracle one in regions with larger quantities of training data and/or more smoothly varying quantiles, and less so in other regions. If we use a method such as a neural network to compute this baseline, typically the baseline interval undercovers on average over the calibration (and test) data, as demonstrated in \citet{Romano2019ConformalizedQR}, for example. 

For each of the existing methods summarized above, this initial interval $\hat{C}_{\textnormal{base}}$ appears in the nested family of sets: specifically, it is recovered by taking $t=0$ for CQR, CQR-r, and CQR-m
(assuming $\alpha_{\textnormal{lo}}=\alpha/2$ and $\alpha_{\textnormal{hi}}=1-\alpha/2$), and by taking $t=\alpha/2$ for DCP. Now, if $\hat{C}_{\textnormal{base}}$ undercovers on the calibration set, conformal prediction leads us to choose a larger value of $\hat{t}$, thus inflating $\hat{C}_{\textnormal{base}}$ in order to achieve a marginal coverage guarantee~\eqref{eqn:marginal_coverage}.

Let us now consider how this inflation behaves. CQR provides an \emph{equal} amount of inflation of $\hat{C}_{\textnormal{base}}(x)$ for every $x$, while 
CQR-r and CQR-m provide a \emph{higher} amount of inflation of $\hat{C}_{\textnormal{base}}(x)$ at values $x$ with higher aleatoric uncertainty (i.e., higher variability in the conditional distribution of $Y \mid X$), and generally this is the case for DCP as well.

In contrast, it would be more efficient to inflate $\hat{C}_{\textnormal{base}}(x)$ primarily at those values $x$ where the initial estimate undercovers---that is, regions with high error in $\hat{q}_{Y \mid X}(x,a)$ as an estimate of the true conditional $a$-quantile for $a=\alpha/2$ and/or $a=1-\alpha/2$. In other words, we would do best to inflate $\hat{C}_{\textnormal{base}}(x)$ more in regions of high \emph{epistemic} uncertainty, which we define as reducible uncertainty (e.g., uncertainty that could have been reduced if we had more data), following the definitions in \citet{senge2014reliable} and \citet{hullermeier2021aleatoric}.

Interestingly, the main empirical conclusion of \citet{sesia2020comparison} is that CQR-r and the related CQR-m, which provide varying amounts of inflation at different values $x$, were unable to provide smaller intervals on average than CQR, whose construction has constant inflation at each $x$. We might be tempted to conclude from this that incorporating local adaptivity into the inflation of CQR is not beneficial. However, an alternative hypothesis explored in this paper is that the issue lies with CQR-r 
 and CQR-m 
inflating the baseline interval proportional to the aleatoric uncertainty, which may be completely different from the epistemic uncertainty; thus, 
we instead conjecture that 
it is important to be locally adaptive to the right source of uncertainty (that is, epistemic rather than aleatoric).

\newtext{\subsection{Additional Related Work}

One of our main proxies for epistemic uncertainty uses the bagging structure of Quantile Regression Forests. \cite{kim2020predictive} and \cite{gupta2022nested} propose using bagging structure for a different purpose: to avoid data splitting for conformal. These papers leverage that each bagged estimator is only trained on a subset of the training data in order to conformalize them by just using out-of-bag residuals, albeit with modified coverage guarantees. \cite{gupta2022nested} adapt this strategy to conformalizing quantile estimators. A characteristic of their algorithm is that the prediction intervals will be smaller when there is consensus among the ensemble of quantile regressors and larger when there is disagreement. This variation among the ensemble members is termed ``instability.''  As we discuss in \cref{sec:heuristics}, we leverage instability as one way to estimate epistemic uncertainty. In this sense, previous work has implicitly taken epistemic uncertainty into account -- but not yet in our split conformal setting. In addition, the inflation of \cite{gupta2022nested}'s intervals when using unstable quantile regressors may also depend on aleatoric uncertainty in subtle ways, while our method more directly separates the two.

Prior conformal methods have incorporated uncertainty estimates. For example, \cite{lei2018distribution} scale the $\hat t$ in \eqref{eqn:split_conformal_simple} by a non-negative multiplicative factor. In that framework, there are many possible choices of scaling factor, including an estimate of the conditional standard deviation and an estimate of the epistemic uncertainty in the conditional mean estimator. A positive of UACQR is that when using quantile regressors the appropriate choice of scaling factor is simply the epistemic uncertainty of the quantile regressors. In contrast, for \cite{lei2018distribution}, the optimal choice may incorporate both aleatoric and epistemic uncertainty. 

\cite{pmlr-v108-izbicki20a} propose a conformal procedure for conditional density estimates to provide prediction sets that not may not be intervals. As with CQR, they estimate aleatoric uncertainty directly and then conformalize the estimates. One can generalize our uncertainty-aware framework to this setting with likely similar benefits when appropriate.}

%% file: Chapters/Uncertainty_aware_CQR.tex
\section{UNCERTAINTY-AWARE CQR\label{sec:UACQR}}

Our proposal is to integrate uncertainty awareness into CQR via the general construction~\eqref{eqn:nested_q_C_hat}, by explicitly incorporating an estimate of epistemic uncertainty into the construction.
We now present two specific proposals that achieve this aim.

\paragraph{UACQR-S} Our first proposed method is similar in construction to the approach of CQR-r and CQR-m, where the baseline quantiles are inflated proportional to a scaling factor. (Here the ``S'' in UACQR-S denotes \emph{scaling}.) We define
\ifdefined\twocolumnstyle
\begin{align*}
    &\hat{q}_{\textnormal{lo}}(x, t) = \hat{q}_{Y \mid X}(x, \alpha_{\textnormal{lo}}) - t\cdot \hat{g}_{\textnormal{lo}}(x), \\
    &\hat{q}_{\textnormal{hi}}(x, t) = \hat{q}_{Y \mid X}(x, \alpha_{\textnormal{hi}}) + t\cdot \hat{g}_{\textnormal{hi}}(x),
\end{align*}
\else
\[
    \hat{q}_{\textnormal{lo}}(x, t) = \hat{q}_{Y \mid X}(x, \alpha_{\textnormal{lo}}) - t\cdot \hat{g}_{\textnormal{lo}}(x), \quad
    \hat{q}_{\textnormal{hi}}(x, t) = \hat{q}_{Y \mid X}(x, \alpha_{\textnormal{hi}}) + t\cdot \hat{g}_{\textnormal{hi}}(x),
\]
\fi
indexed by $t\in\mathbb{R}$, where $\hat{q}_{Y \mid X}(x,\alpha_{\textnormal{lo}})$ 
(respectively, $\hat{q}_{Y \mid X}(x,\alpha_{\textnormal{hi}})$) is some baseline initial estimator of the conditional $\alpha_{\textnormal{lo}}$- (respectively, $\alpha_{\textnormal{hi}}$-) quantile, 
while $\hat{g}_{\textnormal{lo}}(x)$ (respectively, 
$\hat{g}_{\textnormal{hi}}(x)$) estimates the standard deviation of this quantile estimate. (To give a concrete example, 
if we construct $\hat{q}^b_{Y \mid X}(x,a)$'s via bootstrapping subsamples of the training data for each $a=\alpha_{\textnormal{lo}},\alpha_{\textnormal{hi}}$, then $\hat{q}_{Y \mid X}(x,a)$  might be obtained by averaging the $B$ bootstrapped estimates, while $\hat{g}_{\textnormal{lo}}(x)$ and $\hat{g}_{\textnormal{hi}}(x)$ might be computed via the sample standard deviations of the $\hat{q}^b_{Y \mid X}(x,a)$'s at each value of $x$.)

Applying our general construction~\eqref{eqn:nested_q_C_hat} then yields the prediction interval
\ifdefined\twocolumnstyle
\begin{align}\label{eqn:UACQR-S}\hat{C}_n(X_{n+1}) = [&\hat{q}_{Y \mid X}(X_{n+1},\alpha_{\textnormal{lo}})-\hat{t}\cdot \hat{g}_{\textnormal{lo}}(X_{n+1}),\notag\\
&\hat{q}_{Y \mid X}(X_{n+1},\alpha_{\textnormal{hi}})+\hat{t}\cdot \hat{g}_{\textnormal{hi}}(X_{n+1})],\end{align}
\else
\begin{equation}\label{eqn:UACQR-S}\hat{C}_n(X_{n+1}) = [\hat{q}_{Y \mid X}(X_{n+1},\alpha_{\textnormal{lo}})-\hat{t}\cdot \hat{g}_{\textnormal{lo}}(X_{n+1}),\hat{q}_{Y \mid X}(X_{n+1},\alpha_{\textnormal{hi}})+\hat{t}\cdot \hat{g}_{\textnormal{hi}}(X_{n+1})],\end{equation}
\fi
where $\hat{t}$ is computed as in~\eqref{eqn:nested_q_t_hat}.

Examining this construction for UACQR-S, we can observe that CQR-r and CQR-m can be written in the same format---for example, by taking $\hat{g}_{\textnormal{lo}}(x) = \hat{q}_{Y \mid X}(x,0.5)-\hat{q}_{Y \mid X}(x,\alpha_{\textnormal{lo}})$ for CQR-m.
However, for CQR-r and CQR-m the corresponding scaling factors $\hat{g}_{\textnormal{lo}}$ and $\hat{g}_{\textnormal{hi}}$ correspond to the aleatoric uncertainty in our initial quantile regression, while  our particular choice of the scaling factors $\hat{g}_{\textnormal{lo}}$ and $\hat{g}_{\textnormal{hi}}$ is aimed at estimating the epistemic uncertainty, thus providing a more useful local adaptivity 
when inflating the prediction interval.

\paragraph{UACQR-P} Our second variant is UACQR-P, where the ``P'' denotes \emph{percentile}.  
To begin, we compute an ensemble of $B$  estimates of 
$q_{Y \mid X}(x,\alpha_{\textnormal{lo}})$ and of $q_{Y \mid X}(x,\alpha_{\textnormal{hi}})$, denoted by $\hat{q}_{Y \mid X}^b(x,a)$ for each $b \in \{1,\dots,B\}$ and each $a\in\{\alpha_{\textnormal{lo}},\alpha_{\textnormal{hi}}\}$.
(For instance, we may obtain these by running a quantile regression procedure on $B$ different subsets of the training set.)

Let $\hat{q}_{Y \mid X}^{(b)}(x, a)$ refer to the $b$-th order statistic for each $a=\alpha_{\textnormal{lo}},\alpha_{\textnormal{hi}}$, so that the estimates are now sorted, with $\hat{q}_{Y \mid X}^{(1)}(x, a)\leq \dots\leq \hat{q}_{Y \mid X}^{(B)}(x, a)$.
We  also define $\hat{q}_{Y \mid X}^{(0)}(x, a)=-\infty$ and $\hat{q}_{Y \mid X}^{(B+1)}(x, a)=+\infty$.
Then defining
\[\hat{q}_{\textnormal{lo}}(x, t) = \hat{q}_{Y \mid X}^{(B+1-t)}(x, \alpha_{\textnormal{lo}}),\quad
\hat{q}_{\textnormal{hi}}(x, t) = \hat{q}_{Y \mid X}^{(t)}(x, \alpha_{\textnormal{hi}}),\]
over the index set $t\in\mathcal{T}=\{0,1,\dots,B,B+1\}$, applying our general construction~\eqref{eqn:nested_q_C_hat} yields the prediction interval
\begin{equation}\label{eqn:UACQR-P}\hat{C}_n(X_{n+1}) = [\hat{q}^{(B+1-\hat{t})}_{Y \mid X}(X_{n+1},\alpha_{\textnormal{lo}}),\hat{q}^{(\hat{t})}_{Y \mid X}(X_{n+1},\alpha_{\textnormal{hi}})],\end{equation}
where $\hat{t}$ is computed as in~\eqref{eqn:nested_q_t_hat}. 

One useful property of this construction is that it preserves the smoothness properties of the underlying base estimators---in the \cref{apdx:theory}, we show that if each of the bootstrapped estimates $\hat{q}^b_{Y \mid X}(x,a)$ is Lipschitz as a function of $x$, then each order statistic $\hat{q}^{(b)}_{Y \mid X}(x,a)$ is Lipschitz as well.

\subsection{Strategies for estimating epistemic uncertainty} \label{sec:heuristics}
In UACQR-P, epistemic uncertainty is estimated indirectly, via producing an ensemble of quantile estimates $\{\hat{q}^b_{Y \mid X}(x,a)\}_{b=1,\dots,B}$ at each $a=\alpha_{\textnormal{lo}},\alpha_{\textnormal{hi}}$. For UACQR-S, in contrast, we require explicit estimates of epistemic uncertainty, $\hat{g}_{\textnormal{lo}}(x)$ and $\hat{g}_{\textnormal{hi}}(x)$.
Here we outline a few potential strategies for generating these ensembles or these uncertainty estimates.

\paragraph{Bootstrapping/bagging/ensembling.} We can generate a collection $\{\hat{q}^b_{Y \mid X}(x,a)\}_{b=1,\dots,B}$ by training each $\hat{q}^b_{Y \mid X}$ on a different subset of the training data, e.g., by bootstrapping \citep{EfroTibs93} the training set. The empirical distribution of this collection approximates the distribution of $\hat{q}_{Y \mid X}$ under different draws of the training data. For UACQR-S, we can compute
    \[\hat{q}_{Y \mid X}(x,\alpha_{\textnormal{lo}}) = \frac{1}{B}\sum_{b=1}^B \hat{q}^b_{Y \mid X}(x,\alpha_{\textnormal{lo}})\]
    and 
    \[\hat{g}_{\textnormal{lo}}(x) = \left[\frac{1}{B}\sum_{b=1}^B \left(\hat{q}^b_{Y \mid X}(x,\alpha_{\textnormal{lo}}) - \hat{q}_{Y \mid X}(x,\alpha_{\textnormal{lo}})\right)^2\right]^{1/2},\]
and $\hat{q}_{Y \mid X}(x,\alpha_{\textnormal{hi}})$ and $\hat{g}_{\textnormal{hi}}(x)$ are defined analogously. Of course, we might also choose to use a different aggregation procedure, e.g., a median or quantile of the bootstrapped estimates, rather than the mean, which corresponds to bagging \citep{breiman96}. In \cref{apdx:computation}, we discuss how we implement our methods with Quantile Regression Forests \citep{JMLR:v7:meinshausen06a} even with their unique aggregation procedure. \cameraready{For neural networks, performance may improve by training each $\hat{q}_{Y \mid X}^b$ on all training samples, only varying the random initialization of weights for each $b$ \citep{lakshminarayanan2017simple}.}

\paragraph{Epochs of training.} Previous work has indicated that neural networks can  fit ``easy'' regions of the feature space well in early epochs while ``difficult'' regions are not fit until later epochs. For instance, \citet{mangalam2019deep} show that deep neural networks learn examples which are learnable by shallow networks first, suggesting that the number of epochs until a training sample is accurately learned may reflect its epistemic uncertainty. Therefore, a heuristic for epistemic uncertainty is to measure how much a neural network's predictions change across epochs. Here, $\{\hat{q}^b_{Y \mid X}(x,a)\}_{b=1,\dots,B}$ would refer to the model's predictions after each epoch $b$. This heuristic may be more suitable for UACQR-S than UACQR-P, since ensemble members corresponding to later epochs may be higher quality than those of earlier epochs. 
\newtext{\cite{huang2017snapshot} explores using cyclic learning rates to make these ensemble members more exchangeable.}

\paragraph{A parametric approach.} Some quantile regression models may provide a distribution for its estimates under parametric assumptions, such as linear quantile regression \citep{10.1007/978-3-642-57984-4_29} \cameraready{or Bayesian learning approaches which place priors on model parameters \citep{neal2012bayesian}}. Specifically, for UACQR-S, the quantile estimates 
    $\hat{q}_{Y \mid X}(x,\alpha_{\textnormal{lo}})$ and $\hat{q}_{Y \mid X}(x,\alpha_{\textnormal{hi}})$ would be fitted via some parametric model, and the scaling parameters $\hat{g}_{\textnormal{lo}}(x)$ and $\hat{g}_{\textnormal{hi}}(x)$ would then be defined as the standard errors of these quantile estimates, as calculated according to the parametric model.

\subsection{Theoretical guarantee}
\label{sec:theory}
\begin{theorem}
    Assume the data points $(X_1,Y_1),\dots,(X_n,Y_n),(X_{n+1},Y_{n+1})$ are exchangeable. Then the UACQR-P and UACQR-S prediction intervals constructed in~\eqref{eqn:UACQR-P} and~\eqref{eqn:UACQR-S}, respectively, satisfy the marginal coverage guarantee $\mathbb{P}\{Y_{n+1}\in\hat{C}_n(X_{n+1})\}\geq 1-\alpha$.\label{thm:marginal_coverage}
\end{theorem}
\begin{proof}
    This result follows immediately from the properties of conformal prediction \citep{vovk_book,gupta2022nested}, once we observe that UACQR-P and UACQR-S can be formulated as a special case of the nested conformal prediction construction, as detailed in~\eqref{eqn:nested_q_C_hat}.
\end{proof}

\newtext{
\cite{sesia2020comparison} provide an asymptotic guarantee of conditional coverage under consistent quantile estimators for CQR, CQR-r, and CQR-m. This result can directly be extended to UACQR-S. We focus on comparing our methods to baselines under less idealized conditions: finite data and potentially misspecified models.  
}

%% file: Chapters/Simulation_case_study.tex
\section{SIMULATION CASE STUDY}
In this section, we demonstrate our methods on simulated data in which we can highlight the differences between aleatoric and epistemic uncertainty. We generate simulated data from the distribution
\[X \sim \mathrm{Beta}(1.2, 0.8), \quad Y \mid X \sim \mathcal{N}(\sin(X^{-3}), X^4),\]
with sample size $n=100$ (and dimension $p=1$).
We compare our UACQR-P and UACQR-S methods against the existing CQR and CQR-r methods, and repeat the experiment for 150 independent trials. For all four methods, the base quantile regression method is a neural network---we note that this base estimator exhibits substantial overfitting in this simulation. Details of the implementation of all four methods are given in \cref{apdx:methods}. We set $\alpha=0.1$ to create 90\% prediction intervals.

\paragraph{Results.} \cref{fig:seed1} shows the resulting prediction bands constructed by each of the four methods, for one trial of the experiment. 
We also show the oracle prediction interval, given by $[q_{Y \mid X}(x,\alpha/2),q_{Y \mid X}(x,1-\alpha/2)]$ (where these now refer to the \emph{true} conditional quantiles of $Y\mid X$).

\ifdefined\twocolumnstyle
  \begin{figure*}[t]
\else
  \begin{figure}
\fi
    \centering
    \ifdefined\twocolumnstyle
          \includegraphics[width=0.7\textwidth]{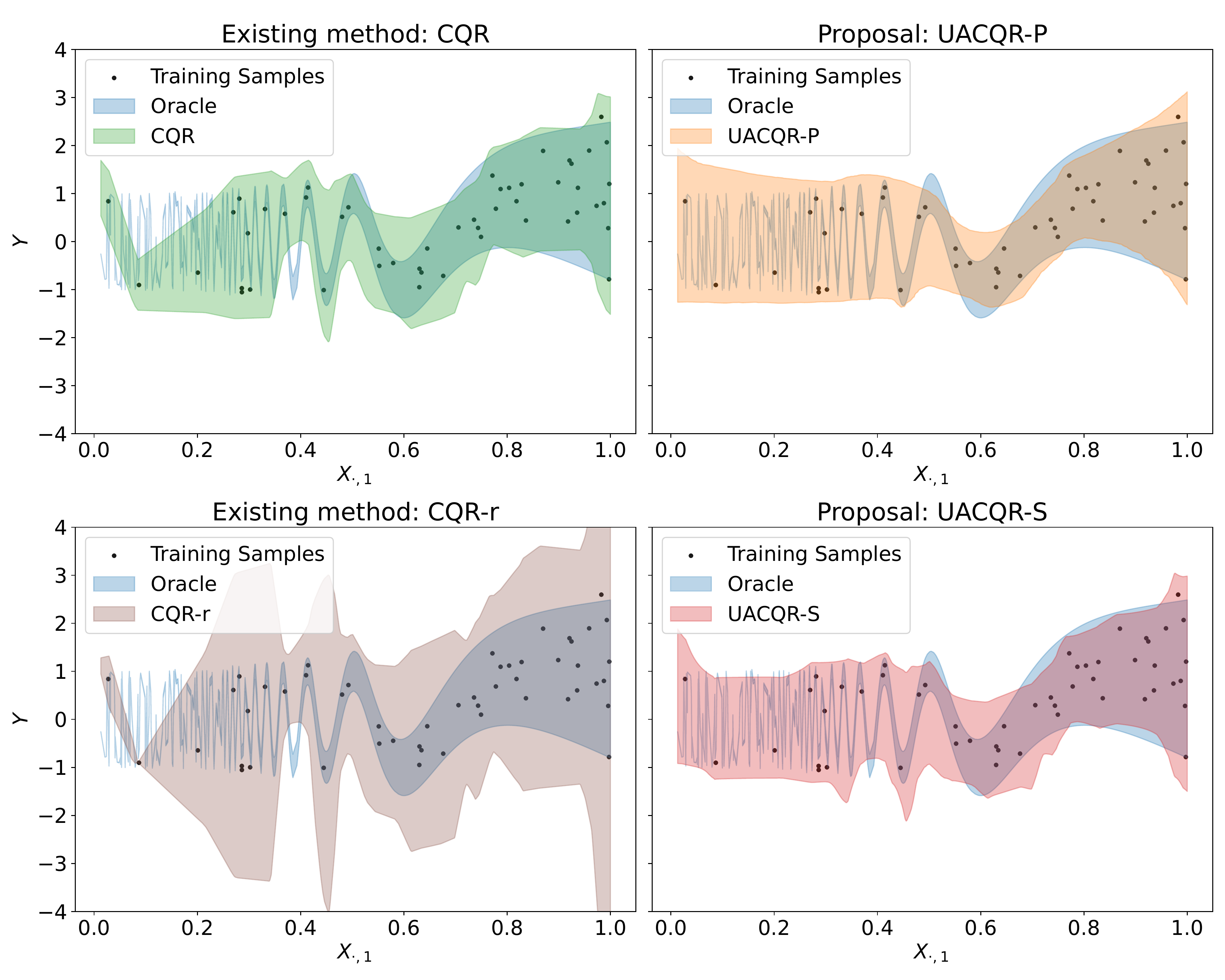}
    \else
          \includegraphics[width=\textwidth]{figures/figure_1_aistats_cr.pdf}
    \fi
    \caption{A comparison of the prediction intervals for existing and proposed methods under one random draw from the data generating process. While existing methods will have low conditional coverage in the high epistemic uncertainty region near $X=0$, our proposals adapt to the toughness of this region. Each conformal procedure is run on the same fitted neural net. For UACQR we use the epoch-based heuristic for epistemic uncertainty. We limit the y-axis to [-4,4] for visual clarity.}
    \label{fig:seed1}
\ifdefined\twocolumnstyle
  \end{figure*}
\else
  \end{figure}
\fi

Examining the oracle prediction interval, in each chart, we can see the distinction between aleatoric and epistemic uncertainty. \emph{Aleatoric uncertainty} is high near $X=1$, since the variance of $Y\mid X$ is higher and thus the oracle prediction interval is relatively wide, and low near $X=0$, where the variance of $Y \mid X$ is low and thus the oracle prediction interval is overly narrow.
\emph{Epistemic uncertainty}, in contrast, is relatively low near $X=1$, since we have many data points that allow our neural net to fit the region well, and much higher near $X=0$, since the Beta distribution for the feature $X$ ensures we have very few training samples
in this region. In addition, the highly non-smooth conditional mean function is particularly challenging to estimate near $X=0$. 

As a consequence, CQR and CQR-r show prediction intervals that are too narrow near $X=0$, which is compensated for by the overly large inflations of the intervals near $X=1$. In contrast, UACQR-P and UACQR-S correctly inflate the prediction intervals more in the high-uncertainty region near $X=0$, and they do not overly inflate the intervals in the more well-informed region near $X=1$.

\begin{figure}[t]
    \centering
    \ifdefined\twocolumnstyle
        \includegraphics[width=0.49\textwidth]{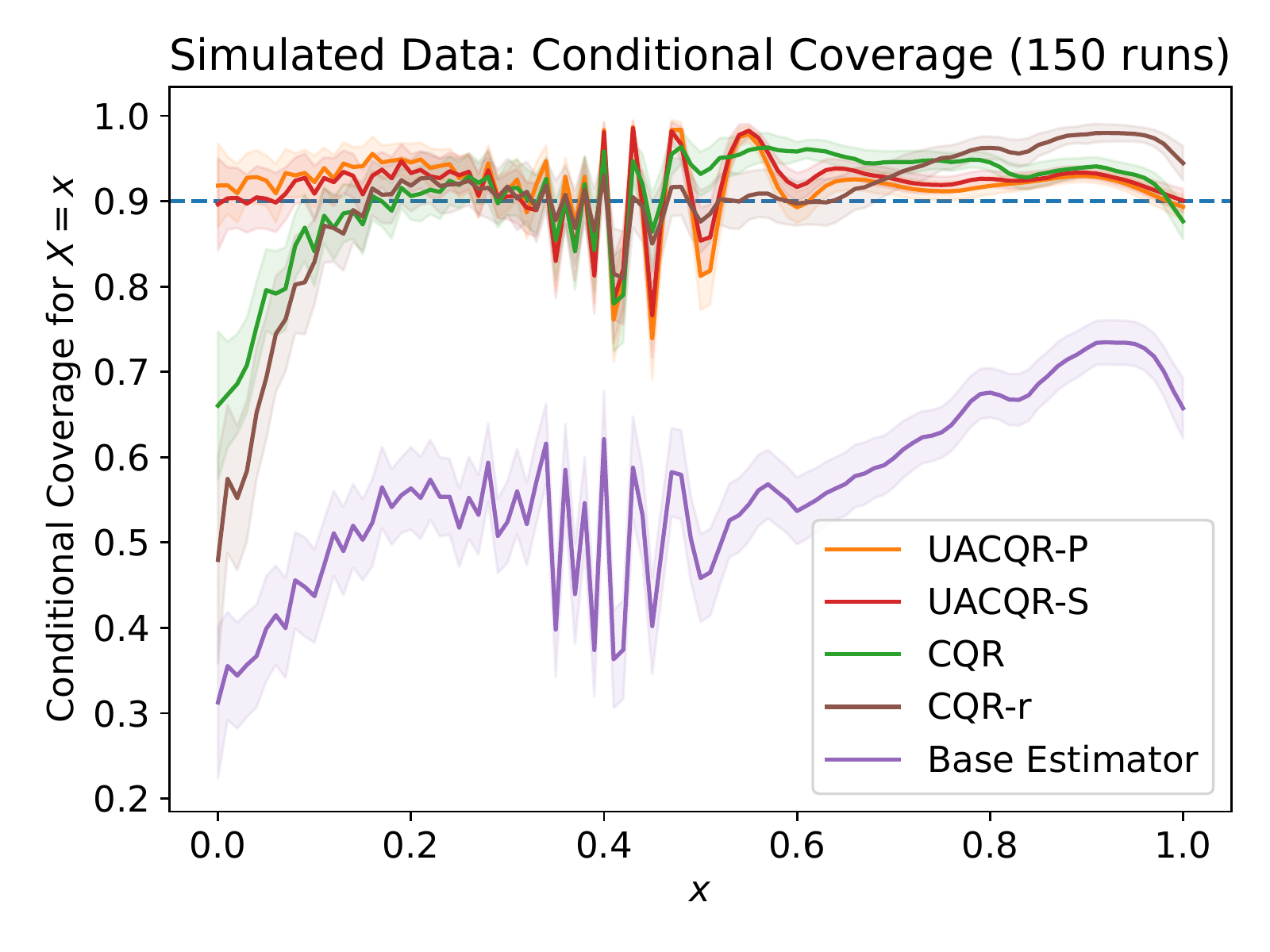}
    \else
        \includegraphics[scale=0.6]{figures/figure_2_aistats_cr.pdf}
    \fi    
    \caption{A comparison of the conditional coverage of existing and proposed methods across 150 random draws, with same setting as \cref{fig:seed1}. All existing methods severely undercover near $X=0$, where epistemic uncertainty is high, while our proposals maintain conditional coverage even in this challenging region.}
    \label{fig:setting_5_average}
    \ifdefined\twocolumnstyle
    \vspace{-10pt}
    \fi
\end{figure}

In \cref{fig:setting_5_average}, we show the results of the experiment averaged over 150 independent trials, where the figure displays the average conditional coverage, $\mathbb{P}\left\{Y_{n+1} \in \hat{C}_n(X_{n+1}) \mid X_{n+1}=x \right\}$, as a function of the test feature value $x\in[0,1]$. While all four methods achieve \emph{marginal} coverage~\eqref{eqn:marginal_coverage}, averaged over $X_{n+1}$ (as guaranteed by the theory), we observe very different empirical trends in the conditional coverage. Specifically, CQR and CQR-r are undercovering in the region of high epistemic uncertainty (near $X=0$), and overcovering in regions of lower epistemic uncertainty (higher values of $X$). In particular, we can also see that the adaptivity of CQR-r to aleatoric uncertainty is counterproductive here, with unnecessarily wide prediction intervals near $X=1$.

In contrast, UACQR-P and UACQR-S maintain relatively even coverage across the range of $X$ values. Of course, these prediction intervals do not fit accurately to the highly non-smooth, high-frequency trends in the conditional mean of $Y \mid X$ near $X=0$, but the wider prediction intervals in this region of high uncertainty compensate appropriately for our inability to estimate this challenging conditional mean. This ability to provide coverage in the high epistemic uncertainty area (near $X=0$) does not come at the expense of overly wide intervals in the low epistemic uncertainty region (larger values of $X$); both of the UACQR methods provide close to 90\% coverage across all values of $X$.

%% file: Chapters/Performance_on_real_world_data_sets.tex
\section{PERFORMANCE ON DATA SETS\label{sec:real-world}}
We measure the performance of our proposals and baselines on data sets from \citet{Romano2019ConformalizedQR} and \citet{sesia2020comparison} in \cref{real_dataset_isl}. 
\ifdefined\twocolumnstyle
  \begin{table*}[t]
\else
  \begin{table}
\fi
  \caption{Real data set results, outperformance bolded, underlined if significant}
  \label{real_dataset_isl}
  \centering
  \begin{tabular}{lllll}
    \toprule
     & \multicolumn{4}{c}{Average Test Interval Score Loss on 20 Runs (Standard Error) }                   \\
    \cmidrule(r){2-5}
    Dataset     & UACQR-P & UACQR-S & CQR & CQR-r \\
    \midrule
\texttt{bike} & \underline{\textbf{1.436}} (0.013) & 1.573 (0.008) & 1.586 (0.008) & 1.564 (0.009) \\
\texttt{bio} & \textbf{1.882} (0.011) & 1.900 (0.010) & 1.900 (0.010) & 1.900 (0.011) \\
\texttt{cbc} & 1.294 (0.026) & 1.276 (0.017) & \textbf{1.274} (0.017) & 1.275 (0.017) \\
\texttt{community} & \textbf{2.103} (0.045) & 2.155 (0.033) & 2.157 (0.034) & 2.156 (0.033) \\
\texttt{concrete} & \textbf{0.834} (0.018) & 0.881 (0.015) & 0.882 (0.015) & 0.864 (0.015) \\
\texttt{forest} & \textbf{2.436} (0.015) & 2.461 (0.010) & 2.462 (0.010) & 2.461 (0.010) \\
\texttt{homes} & \textbf{0.878} (0.012) & 0.916 (0.011) & 0.930 (0.010) & 0.911 (0.011) \\
\texttt{imdb\_wiki} & \underline{\textbf{1.837}} (0.017) & 1.926 (0.010) & 1.928 (0.010) & 1.927 (0.010) \\
\texttt{star} & 0.214 (0.001) & \textbf{0.213} (0.001) & \textbf{0.213} (0.001) & \textbf{0.213} (0.001) \\
    \bottomrule
  \end{tabular}
\ifdefined\twocolumnstyle
  \end{table*}
\else
  \end{table}
\fi
Six of the twelve data sets included in these studies had a response of 0 for over 25\% of samples. This zero-inflation means that methods such as CQR and CQR-r, which inflate the prediction interval symmetrically at the left and right endpoint, are not as well suited for this data. In order to enable a fair comparison with these existing methods, then, we restrict our attention to the remaining six data sets: \texttt{bike} \citep{Dua:2019, bike2}, \texttt{bio} \citep{Dua:2019}, \texttt{community} \citep{Dua:2019, redmond2002data}, \texttt{concrete} \citep{Dua:2019, yeh1998modeling}, \texttt{homes} \citep{data-homes}, and \texttt{star} \citep{data-star}.

\cameraready{We also tested our methods on three new data sets that map images to a continuous response in order to measure the performance of our methods on a wider breadth of tasks. For \texttt{cbc} \citep{alam2019machine}, we estimate the complete blood count from blood smear images. For \texttt{forest} \citep{DeepGlobe18}, from aerial images we estimate the percent of area without forest cover. For \texttt{imdb\_wiki} \citep{Rothe-ICCVW-2015, Rothe-IJCV-2018}, we estimate the year in which celebrity photos from Wikipedia were taken. For all three data sets, we pass the images through a pre-trained ResNet-18 model \citep{he2016deep} to extract 512 continuous features. We then train our quantile regressors using these features.}

We used Quantile Regression Forests for these data sets, since most are tabular and tree-based methods frequently beat neural networks on tabular data \citep{grinsztajn2022tree}. \newtext{We note the average performance across 20 runs in \cref{real_dataset_isl}, bolding outperformance and underlining when statistically significant at the 5\% significance level. Outside simulation settings, we cannot measure conditional coverage, so our metric will instead be the average test interval score loss (defined in \cref{apdx:methods}). The motivation is that interval score loss is a proper scoring rule for prediction intervals, so its expectation is minimized by the true conditional quantiles \citep{gneiting2007strictly}. \cameraready{For users who care about the size of prediction intervals over conditional coverage, we show in \cref{real_dataset_width} that our methods also perform well in terms of average interval width.}
}

Since we use random train-test splits in these experiments, we ensure exchangeability of the data, and therefore all methods will have above 90\% coverage on average. Moreover, we use a randomized conformal procedure, detailed in \cref{apdx:random_conformal}, that ensures all methods average 90\% coverage almost exactly (see \cref{apdx:real_data_extra_experiments}). Our choice of hyperparameter (\texttt{min\_samples\_leaf}) for each method was determined by a cross-validation procedure that minimized the interval score loss for each conformal method. 

In general, it is not necessarily surprising that UACQR-P frequently outperforms UACQR-S here. \newtext{As we observed in \cref{fig:seed1}, prediction intervals from UACQR-P may vary more smoothly with $x$ than those from UACQR-S, as the epistemic uncertainty scaling factors may be noisy.} In addition, UACQR-P is invariant to monotonically increasing transformations of the response since it is a function of ranks, while UACQR-S is not. Therefore, it is possible that the relative performance of UACQR-S could improve under certain transformations. \cameraready{An additional factor is that the machine learning quantile regressors frequently achieve close to 90\% marginal coverage even before conformal prediction (see Table 1 in \citet{Romano2019ConformalizedQR}), leading to a small conformal adjustment ($\hat{t}$) for each of UACQR-S, CQR, and CQR-r. This regime will lead to these three methods performing similarly, regardless of the scaling factor used. In contrast, for our simulation, we chose hyperparameters that encouraged overfitting and yielded far below 90\% coverage without conformal. Accordingly, UACQR-S performs much better relatively in this simulation. }

%% file: Chapters/Discussion_and_limitations.tex
\section{DISCUSSION}
In this work, we have shown that locally adaptive calibration can be beneficial for conformalizing a quantile regression method, as long as the local adaptivity is capturing the correct information---epistemic, rather than aleatoric, uncertainty, or in other words, the variability of the quantile regression estimates, rather than the variability of the response $Y$ itself.

While our methods have demonstrated strong empirical performance on simulations and real-world data sets alike, there are a few caveats. First, the performance of our methods is gated by the ability to estimate epistemic uncertainty well. Thus, for bootstrapping and heuristic approaches alike, our methods may be sensitive to the sample size $n$ and our ensemble size $B$.
\newtext{In addition, our methods are more suitable for high-variance, low-bias quantile regressors, since we estimate epistemic uncertainty by estimating model variance.}
A second potential limitation is that, depending on the specific methods used to compute the initial collection of $B$ estimates, UACQR may present users with a computational-statistical trade-off. \cameraready{An additional limitation is the lack of a theoretical framework for understanding the performance characteristics of UACQR. We suspect that a general theory for this may not be possible but that it is feasible to provide theoretical guarantees in special cases, which may inform a qualitative understanding of the behavior of the method in more general settings.}

On the other hand, if epistemic uncertainty is estimated reasonably well, then (as we see in our experiments), UACQR has the potential to offer prediction intervals with better conditional coverage properties. This demonstrates the benefits of a more locally adaptive approach when implementing conformal prediction, and may offer insights for improving the empirical performance of conformal prediction on problems beyond quantile regression \newtext{-- including classification tasks.}

%% file: Chapters/Acknowledgements.tex
\section{ACKNOWLEDGEMENTS}
R.F.B. was partially supported by the National Science Foundation via grant DMS-2023109, and by the Office of Naval Research via grant N00014-20-1-2337. R.M.W. was partially supported by the National Science Foundation via grants DMS-1930049 and NSF DMS-2023109, and by the Department of Energy via grant DE-AC02-06CH113575.

%% file: Chapters/Fast_heuristic_alternatives.tex
\section{FAST HEURISTIC ALTERNATIVES TO BOOTSTRAPPING}
\label{apdx:computation}
We propose several ways to implement UACQR-P and UACQR-S without needing to refit the base learned model
$B$ times.

\subsection{Quantile Regression Forests}
\begin{figure}[ht!]
\centering
\input{rfqr_speedup_explainer}
\caption{Illustration of a Random Forest, which  serves as a visual aid for the ensuing discussion on existing and proposed methods. \textbf{QRF} outputs the empirical $\alpha_{\textnormal{lo}}, \alpha_{\textnormal{hi}}$ quantiles of the collection of all responses in all the leaves associated with $X_i$ (marked with a star for each tree). \textbf{CQR} then calibrates a constant additive adjustment to these to get $1-\alpha$ coverage. Our sped-up \textbf{UACQR-P}  calculates the empirical $\alpha_{\textnormal{lo}}, \alpha_{\textnormal{hi}}$ quantiles in \textit{each} leaf associated with $X_i$ (marked with a star) and then calibrates how to aggregate these quantile estimates to get $1-\alpha$ coverage.}
\label{rfqr_speedup_graphic}
\end{figure}
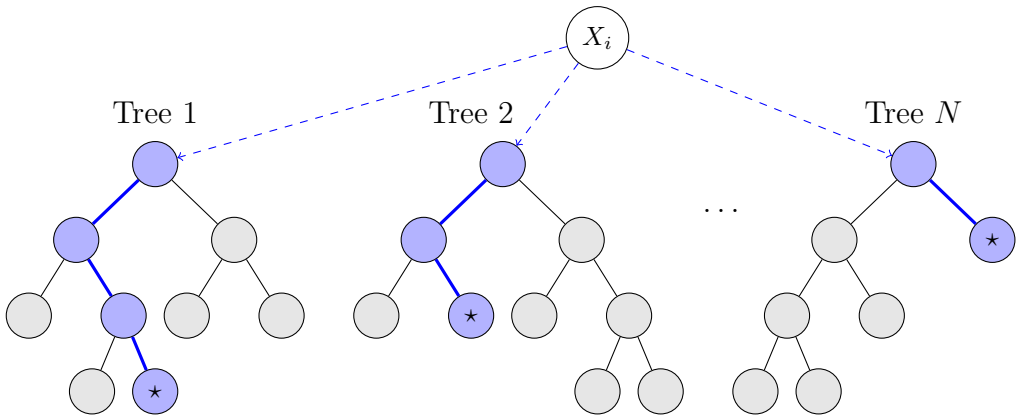

We first describe what happens when one runs Quantile Regression Forests (QRF) \citep{JMLR:v7:meinshausen06a} on a dataset to learn the $a$-quantile. The model is composed of $N$ decision trees, each trained with a bootstrapped sample of points. To make a prediction, we start by feeding the point's covariates $X_{i}$ through each of the decision trees and note which leaves $X_{i}$ resides in. We then take the weighted $a$-quantile of the responses of the training samples in these leaves, where the weights correspond to multiplicity of the training samples in this collection. See \cref{rfqr_speedup_graphic} for reference.

Instead of refitting $B$ full QRFs on bootstrapped training sets, we instead leverage the bootstrapping already done in fitting a single QRF. Namely, we estimate the $a$-quantile in each leaf that $X_i$ resides in and use these $N$ estimates as an ensemble of estimates. Observe that this is equivalent to running UACQR with $B=N$ on QRF models with only one tree in each model. This entails setting $\{\hat{q}^b_{Y \mid X}(x,a)\}_{b=1,\dots,B}$  to be the estimate of the $a$-quantile in each tree.

For UACQR-S,  we can use the standard deviation of this set for $\hat g_{\textnormal{lo}}(x)$ when $a=\alpha_{\textnormal{lo}}$ and similarly for $\hat g_{\textnormal{hi}}(x)$. In our implementations, we set $\hat{q}_{Y \mid X}(x,a)$ to be the output of the full QRF, with $B=N$ trees, for the $a$-quantile. The reason is that this allows the base estimator to be QRF as originally designed, with the heuristic only being used for estimating $\hat g_{\textnormal{lo}}(x), \hat g_{\textnormal{hi}}(x)$.

\subsection{Epoch-based Optimization}
Previous work has indicated that neural networks can  fit ``easy'' regions of the feature space well in early epochs while ``difficult'' regions are not fit until later epochs. For instance, \citet{mangalam2019deep} show that deep neural networks learn examples which are learnable by shallow networks first, suggesting that the number of epochs until a training sample is accurately learned may reflect its epistemic uncertainty. 

Using this idea as a heuristic, one way to approximate epistemic uncertainty is to measure how much a neural network's predictions change across epochs. Here, $\{\hat{q}^b_{Y \mid X}(x,a)\}_{b=1,\dots,B}$ would refer to the model's predictions after each epoch $b$. We can use the standard deviation of this set for $\hat g_{\textnormal{lo}}(x)$ when $a=\alpha_{\textnormal{lo}}$ and similarly for $\hat g_{\textnormal{hi}}(x)$.  

There are two ways to implement this epoch-based heuristic. First, during training, one can store the model's predictions on the held-out samples after each epoch. This may not be feasible if we receive the data in an online fashion and need to train the quantile regressors without access yet to the calibration and test data. Second, one can copy the neural network's parameters after each epoch, and then we can later reload each of these copies during calibration and test to see how the model's predictions for the held-out data changed across epochs. This is more memory intensive, however we did not encounter any problems when using this second approach for our experiments.

One wrinkle, though, is that the user may view the model trained for $B$ epochs as the highest quality estimate. In this case, for UACQR-S we may consider letting $\hat{q}_{Y \mid X}(x,a) = \hat{q}^B_{Y \mid X}(x,a)$ for $a\in\{\alpha_{\textnormal{lo}},\alpha_{\textnormal{hi}}\}$, i.e., the lower and upper bounds of our prediction intervals, pre-conformal, are the fully trained models. This is in fact what we do for our experiments.

It is not quite as easy to privilege the fully-trained model with UACQR-P, and therefore this heuristic may be more suitable for UACQR-S. A subject of future research may be to take weighted percentiles of $\{\hat{q}^b_{Y \mid X}(x,a)\}_{b=1,\dots,B}$ for UACQR-P, where weights increase with $b$. Alternatively, in the bootstrapping literature there is the bias-corrected percentile method \citep{efron1981nonparametric} that, in their setting, incorporates an estimator that uses all of the training samples into the confidence interval construction. \cameraready{In spite of this intuition, our experimental results suggest using UACQR-P over UACQR-S when using the epoch-based heuristic. This outperformance, though, may have depended on using an extensive hyperparameter tuning procedure and always training for at least 100 epochs. In terms of computation and memory, both methods are interchangeable here. See \cref{apdx:method-real_world} for further discussion.}

%% file: rfqr_speedup_explainer.tex

\begin{tikzpicture}[level distance=1.2cm,
level 1/.style={sibling distance=2.5cm},
level 2/.style={sibling distance=1.5cm},
level 3/.style={sibling distance=1cm}, 
every edge/.style={draw},
scale=0.84]
\tikzstyle{input} = [circle, draw, minimum size=0.6cm]
\tikzstyle{node} = [circle, draw, fill=black!10, minimum size=0.6cm]
\tikzstyle{decision} = [circle, draw, fill=black!10, minimum size = 0.6cm]
\tikzstyle{activedecision} = [circle, draw, fill=blue!30, minimum size = 0.6cm]
\tikzstyle{leaf} = [circle, draw, fill=black!10, minimum size = 0.6cm]
\tikzstyle{output} = [circle, draw, fill=blue!30, minimum size = 0.6cm]
\tikzstyle{tree_label} = [above, font=\large]
\node at (0,0) [activedecision] (L14) {} 
    child {node [activedecision] (L13){}
        child {node [leaf]  { }[thin]}
        child {node [activedecision] (L12){}
            child {node [leaf] { }[thin]}
            child {node [output] (L11) { $\star$ }}
        } 
    } 
    child {node [decision] {}
        child {node [leaf] { }}
        child {node [leaf] { }}
    };
\node at (0, 0.5) [tree_label] {Tree 1};
\node at (5.5,0) [activedecision] (L23){}
    child {node [activedecision] (L22){}
        child {node [leaf] { }}
        child {node [output] (L21) { $\star$ }
        }
    }
    child {node [decision] {}
        child {node [leaf] { }}
        child {node [decision] {}
            child {node [leaf] { }
            }child {node [leaf] { }}
        }
    };
\node at (5.0, 0.5) [tree_label] {Tree 2};
\node at (12,0) [activedecision] (L33) {}
    child {node [decision] {}
        child {node [decision] {}
            child {node [leaf] { }}
            child {node [leaf] { }}
        }
        child {node [leaf] { }}
    }
    child {node [output] (L31) { $\star$ }
    edge from parent[very thick, blue]};
\node at (12, 0.5) [tree_label] {Tree $N$};

\node at (9, -0.9) [tree_label] {$\dots$};

\node at (7,2) [input] (X) {$X_i$};

\draw[->,dashed,blue] (X) -- (L14);
\draw[->,dashed,blue] (X) -- (L23);
\draw[->,dashed,blue] (X) -- (L33);

\begin{scope}[very thick, blue]
  \draw (L14) --(L13) -- (L12) -- (L11);
  \draw (L23) -- (L22) -- (L21);
\end{scope}

\end{tikzpicture}

%% file: Chapters/Methodological_details.tex
\section{METHODOLOGICAL DETAILS}
\label{apdx:methods}

All of our experiments were done on a computing cluster, with no extra memory allocations or extra computing nodes requested. We use 150 trials for simulations and 20 trials for the real-world data set experiments. In each trial, we have a new draw of the data (training, calibration, and test) and a different random initialization of the model.

\subsection{Simulation}
\label{sec:nndetails}

We use a neural network with 2 hidden layers, with 100 nodes in each, and ReLU activations for each hidden layer. The neural network outputs three responses, each with the different loss function. The first two are trained with quantile loss, with the target quantiles being 0.05 and 0.95. The third response is trained with squared error loss. We find that adding the squared error loss output dimension, while not theoretically necessary for quantile regression, improves finite sample performance. We use batch normalization \citep{ioffe2015batch} and min-max normalization of both the inputs and the responses. The parameters of the neural network are initialized using PyTorch's default setting. We train the neural network for 1000 epochs with a batch size of 2 and an initial learning rate of 0.001. To make batch normalization more stable in the presence of such a small batch size, we use the running moment estimates after the first epoch, instead of continuing to estimate the moments for each batch. We use a step learning rate scheduler, with a decay rate of $0.999$ every 10 epochs, and the Adam optimizer \citep{kingma2014adam}. These hyperparameter choices were made to encourage overfitting, since if the neural network learned the conditional quantiles well then the benefits of any of the conformal methods would be minimal. For implementing both UACQR methods, we use the epoch-based heuristic to avoid refitting and set $B=999$. For UACQR-S, CQR, and CQR-r, we let the base estimator $\hat{q}_{Y \mid X}(x,a)$ be the fully trained network $\hat{q}^B_{Y \mid X}(x,a)$.

Quantile crossing is a phenomenon in which a quantile regressor may predict a lower value for an upper quantile than for a lower quantile. This tends to occur out-of-sample, when extrapolating. A common post-processing mechanism to address quantile crossing is isotonization: when estimating two quantiles, in the event of crossing, set the upper and lower quantile outputs equal to the average of the two \citep{mammen1991nonparametric, chernozhukov2010quantile}. We ran our experiments with and without this post-processing and received nearly identical results. The likelihood of quantile crossing was reduced in our simulation setting, since the distribution of $X$ was bounded.

\subsection{Real-world data sets\label{apdx:method-real_world}}
We use Quantile Regression Forests
for quantile regression with these data sets. We use default hyperparameter settings, with the exception of \texttt{min\_samples\_leaf}, i.e. the minimum number of samples a leaf can contain. As discussed in \cref{apdx:computation}, the way in which we use Quantile Regression forests is different for UACQR-P from the rest. Since with UACQR-P we calculate the target quantiles in each leaf, there is a heightened need for \texttt{min\_samples\_leaf} to be greater than 1 for our implementation of UACQR-P to have decent estimations in each leaf. We end up choosing this hyperparameter via a cross-validation procedure outlined below. In general, we observe that the optimal \texttt{min\_samples\_leaf} for UACQR-P is usually 5 and 1 for the rest of the methods. For both UACQR methods, we use $B=100$. For UACQR-S, CQR, and CQR-r, we let the base estimator $\hat{q}_{Y \mid X}(x,a)$ be the full QRF, with $B=N=100$ trees, using the standard QRF implementation.

To ensure exchangeability, we sample the training, calibration, and test sets without replacement from the full data sets. We use  40\% of the points for training, 40\% for calibration, and 20\% for testing. For computability, for the \texttt{imdb\_wiki}, \texttt{bio}, and \texttt{homes} data sets, we use only use 25\%, 25\%, and 50\%, respectively, of the full data sets in any run. For clarity, with \texttt{bio} in each run 10\% of all data points are used for training, 10\% for calibration, 5\% for testing, and 75\% not at all.

\newtext{To select hyperparameters, once we do our initial train/calibration/test splits, we further divide the training set again with a 40\%/40\%/20\% breakdown for the purposes of cross-validation. We then perform each of the conformal procedures using this sub-splitting of the training data with the various hyperparameter settings. We note the hyperparameter setting for each method that minimizes interval score loss on the 20\% sub-split of the training data. We then use these settings for each method on the initial train/calibration/test splits to measure final performance. In short, we selected hyperparameters without using oracle knowledge or data-snooping, ensuring realistic results.} 

\cameraready{For the \texttt{cbc} data set, we increase the number of samples by also including horizontal and vertical flips of the images, increasing the total number of sample from 360 to 1,080. For the \texttt{imdb\_wiki} data set, we note that we only use the Wikipedia images to reduce the number of samples to a more manageable size. In addition, we only use images taken before 2005, since otherwise the distribution of when photos were taken would skew heavily toward more recent years.}

As done in \citet{Romano2019ConformalizedQR}, we divide the responses by the mean absolute value response of the training set. This ensures that the rows in \cref{real_dataset_isl} each have similar magnitude.

We do not need to worry about quantile crossing for the experiments with Quantile Regression Forests, since each quantile estimate corresponds to the sample quantile of the same set of samples. We undo quantile crossing for our neural network experiment on real-world data.

In all experiments, including the additional ones in \cref{apdx:real_data_extra_experiments}, for UACQR-S we calculate $\hat g_{\textnormal{lo}}(x),\hat g_{\textnormal{hi}}(x)$ using standard deviations, as described in \cref{sec:heuristics}. However, we have observed that UACQR-S's performance on these real data experiments often improves if we use interquartile range instead, likely due to the metric's robustness to outliers. 

\newtext{\subsection{Evaluation Metrics}

\paragraph{Interval Score Loss} For real-world data sets, we mainly evaluate our methods with interval score loss:
\begin{equation}
    \begin{split}
        \ell^{\text{is}} (y,\hat{q}_{\textnormal{lo}}(x,\hat{t}), \hat{q}_{\textnormal{hi}}(x,\hat{t}); \alpha) &= \hat{q}_{\textnormal{hi}}(x,\hat{t}) -\hat{q}_{\textnormal{lo}}(x,\hat{t}) \\
        &+ 
        \frac{2}{\alpha} \cdot (\hat{q}_{\textnormal{lo}}(x,\hat{t})-y)\mathbf{1}_{y<\hat{q}_{\textnormal{lo}}(x,\hat{t})} \\
        &+ 
        \frac{2}{\alpha} \cdot (y-\hat{q}_{\textnormal{hi}}(x,\hat{t}))\mathbf{1}_{y>\hat{q}_{\textnormal{hi}}(x,\hat{t})}
    \end{split}
\end{equation}

The first two terms can be viewed as the interval width, and the remaining terms can be viewed as penalties for not covering by the added distance needed to have covered. Therefore, by minimizing interval score loss, we are balancing the goal of minimizing interval widths with the goal of not undercovering. 

As mentioned in \cref{sec:real-world}, interval score loss is a proper scoring rule \citep{gneiting2007strictly}, so it is minimized in expectation by the true $\frac{\alpha}{2}, 1-\frac{\alpha}{2}$ quantiles. In fact, interval score loss is proportional to the sum of pinball losses for the $\frac{\alpha}{2}, 1-\frac{\alpha}{2}$ quantiles. Therefore, it is ideally used for measuring the performance of $1-\alpha$ prediction intervals that are intended to be \emph{symmetric}, with equal probability of miscovering above and below. This matches our use case.

\paragraph{Additional metrics}
\citet{feldman2021improving} propose additional metrics that may be useful for measuring conditional coverage properties. Chief among their suggestions is the dependence between a coverage indicator variable $\mathbf{1}_{Y_i \in \hat{C}_n(X_i)}$ and interval width $\hat{q}_{\textnormal{hi}}(X_i,\hat{t}) -\hat{q}_{\textnormal{lo}}(X_i,\hat{t})$, usually as measured by the correlation between the two random variables. If there is dependence, then we may view this as a violation of conditional coverage, since a necessary condition of conditional coverage is $\mathbf{1}_{Y_i \in \hat{C}_n(X_i)}$ being independent of \emph{all} measurable functions of $X_i$. With this in mind, we see that achieving zero correlation on their metric is only a necessary but not sufficient condition for conditional coverage, as even degenerate constructions can achieve zero correlation, such as any prediction interval with constant width. \citet{feldman2021improving} provide additional metrics, but each is also susceptible to a degenerate construction scoring well. In contrast, it may be very difficult or impossible to minimize interval score loss in expectation, depending on regularity conditions, without using the true $\frac{\alpha}{2}, 1-\frac{\alpha}{2}$ conditional quantiles.
}

%% file: Chapters/Randomized_cutoffs.tex
\section{RANDOMIZED CUTOFFS FOR EXACT COVERAGE}\label{apdx:random_conformal}
The coverage guarantee obtained by methods within the conformal prediction framework is generally written as a lower bound, i.e., $\mathbb{P}\{Y_{n+1}\in\hat{C}_n(X_{n+1})\}\geq 1-\alpha$. Examining the construction for the nested split conformal case specifically, given in~\eqref{eq:nested_that} and~\eqref{eqn:nested_q_C_hat}, we can see that the true probability of coverage might be higher than $1-\alpha$, for two reasons: first, due to rounding error (if $(1-\alpha)(n_1+1)$ is not an integer), and second, due to ties among the conformity scores, which in this case are given by the values 
\[t_i = \inf\{t\in\mathcal{T}: Y_i \in\hat{C}_{n_0,t}(X_i)\}.\]
These issues may be problematic when comparing the empirical performance of various methods---e.g., if we see one method has wider prediction intervals than another, is this a genuine difference or might it simply be due to overcoverage arising from rounding error or from ties?  For this reason we use a modified version of the procedure for our real data experiments, as detailed next.

To alleviate this issue of overcoverage, \citet{vovk_book}
proposes a randomization strategy, referred to as ``smoothed conformal predictors'' in that work. Here we show calculations for how to specialize \citet{vovk_book}'s randomization strategy to the specific setting of split conformal via nested sets, as constructed in~\eqref{eqn:nested_q_C_hat}.

Let $k = \lceil (1-\alpha)(n_1 + 1) \rceil$, and let $\delta = k - (1-\alpha)(n_1+1)$, capturing the rounding error.
Let $t_i$ be defined as above for each $i=n_0+1,\dots,n$ in the calibration set, and define the order statistics $t_{(1)}\leq \dots \leq t_{(n_1)}$ of this list. Define
\[T_0 = \#\{i \leq k-1 : t_{(i)} = t_{(k-1)}\}\textnormal{ and }T_1 = \#\{i \geq k : t_{(i)} = t_{(k)}\},\]
which capture information about the number of ties for $(k-1)$st or $k$th place.
Then we split into two cases: if $t_{(k-1)}<t_{(k)}$, then define
\[\hat{C}_n(X_{n+1}) = \begin{cases}
    (\hat{q}_{\textnormal{lo}}(X_{n+1},t_{(k-1)}),\hat{q}_{\textnormal{hi}}(X_{n+1},t_{(k-1)})), & \textnormal{ with probability }\frac{\delta }{T_0+1},\\
    [\hat{q}_{\textnormal{lo}}(X_{n+1},t_{(k-1)}),\hat{q}_{\textnormal{hi}}(X_{n+1},t_{(k-1)})], & \textnormal{ with probability }\frac{\delta T_0}{T_0+1},\\
    (\hat{q}_{\textnormal{lo}}(X_{n+1},t_{(k)}),\hat{q}_{\textnormal{hi}}(X_{n+1},t_{(k)})), & \textnormal{ with probability }\frac{(1-\delta)T_1}{T_1+1},\\
    [\hat{q}_{\textnormal{lo}}(X_{n+1},t_{(k)}),\hat{q}_{\textnormal{hi}}(X_{n+1},t_{(k)})], & \textnormal{ with probability }\frac{1-\delta}{T_1+1},
\end{cases}\]
while if instead $t_{(k-1)}=t_{(k)}$, then define

\[\hat{C}_n(X_{n+1}) = \begin{cases}
    (\hat{q}_{\textnormal{lo}}(X_{n+1},t_{(k)}),\hat{q}_{\textnormal{hi}}(X_{n+1},t_{(k)})), & \textnormal{ with probability }\frac{T_1+\delta}{T_0+T_1+1},\\
    [\hat{q}_{\textnormal{lo}}(X_{n+1},t_{(k)}),\hat{q}_{\textnormal{hi}}(X_{n+1},t_{(k)})], & \textnormal{ with probability }\frac{T_0+1-\delta}{T_0+T_1+1}.
\end{cases}\]

Note that we might have $k = n_1+1$ (i.e., the order statistic $t_{(k)}$ is not defined, or rather, is taken to be $+\infty$). This happens in the case where $\alpha(n_1+1) < 1$ . In this case, we take $t_{(n_1+1)} := +\infty$ and all the above definitions are still valid. (Observe that this scenario would fall into the first case, $t_{(k-1)} < t_{(k)}$, since $t_{(k-1)} = t_{(n_1)}$ is finite while $t_{(k)}$ is set to be $+\infty$.)

%% file: Chapters/Lipschitz_property.tex
\section{LIPSCHITZ PROPERTY FOR UACQR-P}\label{apdx:theory}

As mentioned earlier, the UACQR-P construction preserves the smoothness properties of the underlying base estimators. 
Here we will verify that if each of the bootstrapped estimates $\hat{q}^b_{Y \mid X}(x,a)$ is $L$-Lipschitz (as a function of $x$), then this property is preserved by the function $\hat{q}^{(b)}_{Y \mid X}(x,a)$, for each $b$. In particular, the left- and right-endpoint functions of the prediction band $\hat{C}_n = \{\hat{C}_n(x)\}$ will then be $L$-Lipschitz as well. The following basic result verifies this claim.

\begin{proposition}
Let $f_1,\dots,f_B:\mathcal{X}\rightarrow\mathbb{R}$, where $\mathcal{X}$ is a normed space. For any $b\in\{1,\dots,B\}$, define the function $f_{(b)}:\mathcal{X}\rightarrow\mathbb{R}$ as \[f_{(b)}(x) = \inf\left\{t\in\mathbb{R} : \sum_{i=1}^B \mathbf{1}_{f_i(x)\leq t} \geq b\right\}.\]
(In other words, for each $x$, $f_{(1)}(x)\leq \dots \leq f_{(B)}(x)$ are the order statistics of $f_1(x),\dots,f_B(x)$.)

Then, if $f_b$ is $L$-Lipschitz for all $b$, it holds that $f_{(b)}$ is $L$-Lipschitz for all $b$.
\end{proposition}
\begin{proof}
Suppose $f_{(b)}$ is not $L$-Lipschitz. Then there must exist some $x_0,x_1\in\mathcal{X}$ with $|f_{(b)}(x_0) - f_{(b)}(x_1)|>L\|x_0-x_1\|$. 
Without loss of generality, we can assume $f_{(b)}(x_1)>f_{(b)}(x_0)$. 

By definition of $f_{(b)}(x_0)$ as the $b$th order statistic of $f_1(x_0),\dots,f_B(x_0)$, we have
\[\sum_{i=1}^B \mathbf{1}_{f_i(x_0)\leq f_{(b)}(x_0)} \geq b.\]
Similarly, by definition of $f_{(b)}(x_1)$ as the $b$th order statistic of $f_1(x_1),\dots,f_B(x_1)$, we have
\[\sum_{i=1}^B \mathbf{1}_{f_i(x_1)\geq f_{(b)}(x_1) }\geq B-b+1.\]
Combining these two facts, we must have at least one $i\in\{1,\dots,B\}$ for which it holds that
\[f_i(x_0)\leq f_{(b)}(x_0) \textnormal{ and }f_i(x_1)\geq f_{(b)}(x_1)> f_{(b)}(x_0)+ L\|x_0-x_1\| .\]
This is a contradiction, since $f_i$ is $L$-Lipschitz.
\end{proof}

%% file: Chapters/Additional_experiments.tex
\section{ADDITIONAL EXPERIMENTS}
\subsection{Simulation}
\begin{figure}[h]
    \centering
    \includegraphics[scale=0.6]{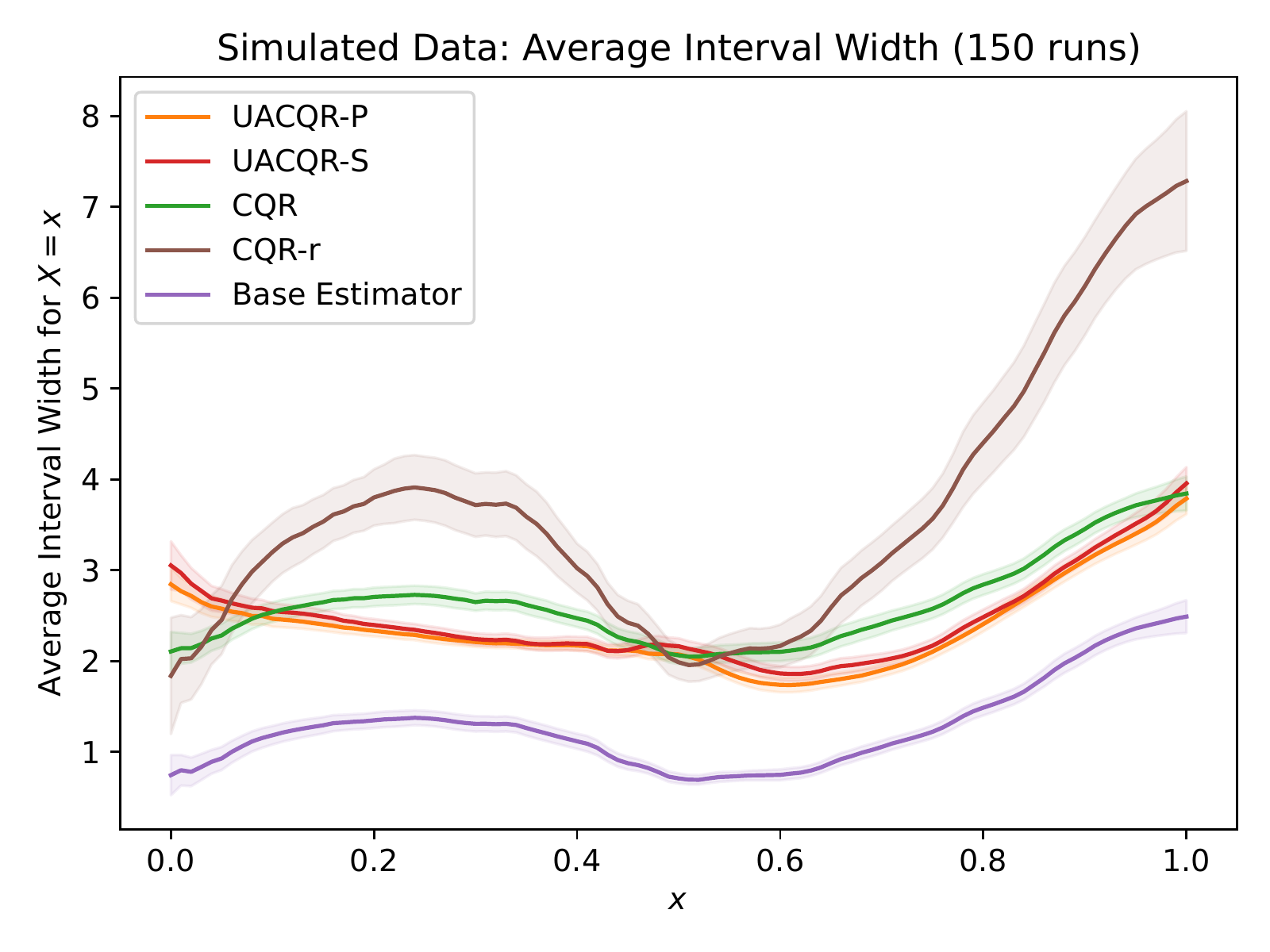}
    \caption{In the same setting as \cref{fig:setting_5_average}, we see that the two UACQR methods have on average shorter interval widths than CQR and CQR-r for most values of $X$. When epistemic uncertainty is high ($X=0$), our methods have wider intervals, allowing them to achieve conditional coverage in this region.}
\label{fig:setting_5_width}
\end{figure}

In \cref{fig:setting_5_average} we saw that UACQR methods are able to maintain conditional coverage near $X=0$, where epistemic uncertainty was high, without overcovering significantly when epistemic uncertainty was low, near $X=1$. In \cref{fig:setting_5_width}, we see that this adaptivity usually yields more precise intervals on average. Near $X=0$, the UACQR methods provide wider intervals than the existing methods in order to achieve conditional coverage. This does not yield wider intervals on average everywhere else, though, with UACQR methods providing intervals as narrow as other methods when epistemic uncertainty is low. This further demonstrates the locally adaptive properties of our methods in this simulation setting.

\subsection{Real-world data sets}\label{apdx:real_data_extra_experiments}
In this section, we test several alternative implementations of our real data experiments, to verify that our results persist across different choices made in the implementation. We also verify the marginal coverage properties, as guaranteed by the theory and the construction of our train/calibration/test splits. 

\newtext{As before, we bold outperformance and underline when statistically significant. Our measure for statistical significance for each data set is a two-sample t-test between the best performing method and the second best performing method, with no multiplicity corrections.}

\newtext{In \cref{real_dataset_cov}, we provide the average test coverage for the experimental results already provided in \cref{sec:real-world}. We confirm that we have the desired marginal coverage of 90\%, with slight variations due only to using 20 runs.

\begin{table}[h]
  \caption{Real data results on 20 runs: marginal coverage}
  \label{real_dataset_cov}
  \centering
  \begin{tabular}{lllll}
    \toprule
     & \multicolumn{4}{c}{Average Test Coverage (Standard Error)}                   \\
    \cmidrule(r){2-5}
    Dataset     & UACQR-P & UACQR-S & CQR & CQR-r \\
    \midrule
\texttt{bike} & 0.904 (0.002) & 0.900 (0.002) & 0.900 (0.002) & 0.900 (0.002) \\
\texttt{bio} & 0.900 (0.001) & 0.900 (0.002) & 0.901 (0.002) & 0.900 (0.002) \\
\texttt{cbc} & 0.896 (0.006) & 0.897 (0.006) & 0.893 (0.006) & 0.894 (0.006) \\
\texttt{community} & 0.905 (0.005) & 0.902 (0.003) & 0.901 (0.003) & 0.899 (0.003) \\
\texttt{concrete} & 0.902 (0.006) & 0.896 (0.006) & 0.893 (0.006) & 0.896 (0.005) \\
\texttt{forest} & 0.902 (0.003) & 0.902 (0.003) & 0.902 (0.003) & 0.902 (0.003) \\
\texttt{homes} & 0.901 (0.002) & 0.901 (0.001) & 0.901 (0.001) & 0.902 (0.002) \\
\texttt{imdb\_wiki} & 0.897 (0.005) & 0.896 (0.005) & 0.893 (0.005) & 0.894 (0.005) \\
\texttt{star} & 0.898 (0.003) & 0.898 (0.004) & 0.898 (0.004) & 0.898 (0.004) \\
    \bottomrule
  \end{tabular}
\end{table}}

\cameraready{In \cref{real_dataset_width}, we see that our methods also usually provide more precise intervals on average, as measured by average test interval width. The one exception is the \texttt{star} data set. We see that these instances of outperformance for UACQR-P are statistically significant in seven of the eight cases.

\begin{table}[h]
  \caption{Real data results on 20 runs: interval width}
  \label{real_dataset_width}
  \centering
  \begin{tabular}{lllll}
    \toprule
     & \multicolumn{4}{c}{Average Test Interval Width (Standard Error)}                   \\
    \cmidrule(r){2-5}
    Dataset     & UACQR-P & UACQR-S & CQR & CQR-r \\
    \midrule
\texttt{bike} & \underline{\textbf{1.184}} (0.007) & 1.456 (0.006) & 1.484 (0.006) & 1.440 (0.006) \\
\texttt{bio} & \underline{\textbf{1.530}} (0.007) & 1.654 (0.004) & 1.656 (0.004) & 1.653 (0.004) \\
\texttt{cbc} & \underline{\textbf{0.827}} (0.013) & 1.004 (0.012) & 0.998 (0.013) & 1.002 (0.012) \\
\texttt{community} & \underline{\textbf{1.487}} (0.028) & 1.774 (0.021) & 1.779 (0.021) & 1.774 (0.022) \\
\texttt{concrete} & \textbf{0.678} (0.015) & 0.734 (0.009) & 0.732 (0.008) & 0.711 (0.008) \\
\texttt{forest} & \underline{\textbf{2.269}} (0.012) & 2.361 (0.008) & 2.363 (0.008) & 2.363 (0.008) \\
\texttt{homes} & \underline{\textbf{0.633}} (0.005) & 0.699 (0.004) & 0.758 (0.004) & 0.714 (0.005) \\
\texttt{imdb\_wiki} & \underline{\textbf{1.558}} (0.018) & 1.799 (0.011) & 1.796 (0.011) & 1.798 (0.011) \\
\texttt{star} & 0.176 (0.001) & 0.176 (0.001) & \textbf{0.175} (0.001) & 0.176 (0.002) \\
    \bottomrule
  \end{tabular}
\end{table}}

\newtext{In \cref{real_dataset_oqr}, we measure performance again in this same setting as \cref{sec:real-world}, but this time we use the \citet{feldman2021improving} metric to measure conditional coverage properties. Specifically, as discussed in \cref{apdx:methods}, we provide the absolute value of the correlation on test data between a coverage indicator variable $\mathbf{1}_{Y_i \in \hat{C}_n(X_i)}$ and interval width $\hat{q}_{\textnormal{hi}}(X_i,\hat{t}) -\hat{q}_{\textnormal{lo}}(X_i,\hat{t})$. Our cross-validation procedure on this experiment chooses hyperparameters that minimize this metric, as opposed to interval score loss. We can see that UACQR-P provides the best results on four of the nine data sets, including in a significant way on one of them. UACQR-S and CQR-r each perform uniquely the best on two data sets, and they tie for best on the final data set. Observe that most of these differences are not statistically significant. As we discussed in \cref{apdx:methods}, this metric may be minimized even under degenerate constructions. Therefore, we interpret these results as still underscoring the strong conditional coverage properties of UACQR-P, by performing better or similarly well on this metric as any other while doing better almost always on the more discriminating interval score loss metric.

\begin{table}[h]
  \caption{Real data results on 20 runs on \citet{feldman2021improving}'s metric}
  \label{real_dataset_oqr}
  \centering
  \begin{tabular}{lllll}
    \toprule
     & \multicolumn{4}{c}{Correlation between Coverage and Width (Standard Error)}                   \\
    \cmidrule(r){2-5}
    Dataset     & UACQR-P & UACQR-S & CQR & CQR-r \\
    \midrule
\texttt{bike} & \underline{\textbf{0.038}} (0.005) & 0.130 (0.008) & 0.217 (0.007) & 0.119 (0.007) \\
\texttt{bio} & 0.028 (0.004) & 0.029 (0.005) & 0.027 (0.004) & \textbf{0.024} (0.004) \\
\texttt{cbc} & \textbf{0.059} (0.012) & 0.072 (0.009) & 0.062 (0.011) & 0.073 (0.012) \\
\texttt{community} & 0.063 (0.009) & \textbf{0.062} (0.009) & 0.066 (0.012) & 0.063 (0.008) \\
\texttt{concrete} & \textbf{0.105} (0.014) & 0.113 (0.014) & 0.197 (0.016) & 0.112 (0.014) \\
\texttt{forest} & 0.054 (0.007) & \textbf{0.047} (0.007) & 0.051 (0.008) & \textbf{0.047} (0.007) \\
\texttt{homes} & 0.029 (0.005) & 0.027 (0.004) & 0.096 (0.004) & \textbf{0.019} (0.004) \\
\texttt{imdb\_wiki} & 0.079 (0.012) & \textbf{0.073} (0.015) & 0.078 (0.014) & 0.078 (0.014) \\
\texttt{star} & \textbf{0.125} (0.011) & 0.132 (0.014) & 0.128 (0.013) & 0.132 (0.015) \\
    \bottomrule
  \end{tabular}
\end{table}}

In \cref{real_dataset_avg_length_log}, we show results after taking a log transformation (specifically, replacing $Y_i$ with $\log(1+Y_i)$ for all data points, to allow for zero values). We do this instead of the usual mean normalization, so the units in \cref{real_dataset_avg_length_log} are different from other tables. All other details of the implementation remain the same. We observe that UACQR-P continues to show the best performance in terms of interval score loss on almost all data sets, with the exception of \texttt{star} and \texttt{imdb\_wiki}.

\begin{table}[h]
  \caption{Real data results on 20 runs with log transformation.}
  \label{real_dataset_avg_length_log}
  \centering
  \begin{tabular}{lllll}
    \toprule
     & \multicolumn{4}{c}{Average Test Interval Score Loss (Standard Error)}                   \\
    \cmidrule(r){2-5}
    Dataset     & UACQR-P & UACQR-S & CQR & CQR-r \\
    \midrule
\texttt{bike} & \underline{\textbf{2.048}} (0.012) & 2.212 (0.009) & 2.224 (0.009) & 2.201 (0.01) \\
\texttt{bio} & \underline{\textbf{1.843}} (0.014) & 1.88 (0.008) & 1.879 (0.008) & 1.879 (0.008) \\
\texttt{cbc} & \textbf{1.582} (0.051) & 1.602 (0.034) & 1.599 (0.034) & 1.598 (0.033) \\
\texttt{community} & \textbf{0.378} (0.006) & 0.379 (0.005) & 0.379 (0.005) & 0.379 (0.005) \\
\texttt{concrete} & \textbf{0.984} (0.019) & 1.027 (0.024) & 1.035 (0.021) & 1.021 (0.025) \\
\texttt{forest} & \underline{\textbf{0.649}} (0.002) & 0.656 (0.001) & 0.656 (0.001) & 0.656 (0.001) \\
\texttt{homes} & \underline{\textbf{0.829}} (0.005) & 0.851 (0.005) & 0.855 (0.004) & 0.846 (0.005) \\
\texttt{imdb\_wiki} & 3.409 (0.061) & 3.393 (0.032) & 3.394 (0.032) & \textbf{3.392} (0.032) \\
\texttt{star} & 0.212 (0.001) & \textbf{0.211} (0.001) & \textbf{0.211} (0.001) & \textbf{0.211} (0.001) \\
    \bottomrule
  \end{tabular}
\end{table}

In \cref{real_dataset_avg_length_noRandom}, we show the results after running the original experiment without the log transformation, except that we do not use randomization for handling ties/rounding (as detailed in \cref{apdx:random_conformal}); instead, we use the original definition of the split nested conformal method given in~\eqref{eqn:nested_q_C_hat} earlier. Here we see that the UACQR-P method is generally producing the lowest interval score loss, with the exception of the \texttt{star} and \texttt{cbc} data sets, same as with \cref{real_dataset_isl}.

\begin{table}[h]
  \caption{Real data results on 20 runs with no randomized cutoffs.}
  \label{real_dataset_avg_length_noRandom}
  \centering
  \begin{tabular}{lllll}
    \toprule
     & \multicolumn{4}{c}{Average Test Interval Score Loss (Standard Error)}                   \\
    \cmidrule(r){2-5}
    Dataset     & UACQR-P & UACQR-S & CQR & CQR-r \\
    \midrule
\texttt{bike} & \underline{\textbf{1.447}} (0.013) & 1.573 (0.008) & 1.586 (0.008) & 1.564 (0.009) \\
\texttt{bio} & \textbf{1.880} (0.012) & 1.900 (0.010) & 1.900 (0.010) & 1.900 (0.011) \\
\texttt{cbc} & 1.278 (0.023) & 1.275 (0.016) & \textbf{1.273} (0.016) & \textbf{1.273} (0.016) \\
\texttt{community} & \textbf{2.112} (0.042) & 2.156 (0.033) & 2.157 (0.034) & 2.156 (0.033) \\
\texttt{concrete} & \textbf{0.842} (0.019) & 0.882 (0.015) & 0.881 (0.015) & 0.864 (0.015) \\
\texttt{forest} & \textbf{2.439} (0.014) & 2.461 (0.010) & 2.462 (0.010) & 2.461 (0.010) \\
\texttt{homes} & \textbf{0.880} (0.012) & 0.916 (0.011) & 0.930 (0.010) & 0.911 (0.011) \\
\texttt{imdb\_wiki} & \underline{\textbf{1.841}} (0.018) & 1.926 (0.010) & 1.927 (0.010) & 1.926 (0.010) \\
\texttt{star} & 0.214 (0.001) & \textbf{0.213} (0.001) & \textbf{0.213} (0.001) & \textbf{0.213} (0.001) \\
    \bottomrule
  \end{tabular}
\end{table}

In \cref{real_dataset_avg_length_neuralnet}, we show the results when the methods are run using a neural network, rather than Quantile Regression Forests, as the base algorithm. The model architecture is similar to what is described in \cref{sec:nndetails}, except that we used an extensive hyperparameter grid search over number of epochs, learning rate, batch size, weight decay penalty, dropout rate \citep{srivastava2014dropout}, and size of the hidden layers. For each method, we report the average test interval score loss (across 20 trials) after choosing for each trial the best hyperparameter choices among 25 randomly drawn options using our cross-validation procedure. Again, we see that the UACQR-P method generally provides the lowest interval score loss.

\begin{table}[h]
  \caption{Real data results on 20 runs with neural network as base algorithm.}
  \label{real_dataset_avg_length_neuralnet}
  \centering
  \begin{tabular}{lllll}
    \toprule
     & \multicolumn{4}{c}{Average Test Interval Score Loss (Standard Error)}                   \\
    \cmidrule(r){2-5}
    Dataset     & UACQR-P & UACQR-S & CQR & CQR-r \\
    \midrule
\texttt{bike} & \underline{\textbf{1.204}} (0.015) & 1.301 (0.018) & 1.322 (0.020) & 1.317 (0.018) \\
\texttt{bio} & \underline{\textbf{2.020}} (0.016) & 2.073 (0.015) & 2.076 (0.016) & 2.074 (0.017) \\
\texttt{cbc} & 1.291 (0.040) & 1.293 (0.027) & 1.265 (0.020) & \textbf{1.256} (0.021) \\
\texttt{community} & \underline{\textbf{2.372}} (0.050) & 2.546 (0.054) & 2.573 (0.044) & 2.598 (0.053) \\
\texttt{concrete} & \textbf{1.832} (0.028) & \textbf{1.832} (0.024) & 1.839 (0.025) & 1.842 (0.026) \\
\texttt{forest} & \underline{\textbf{2.639}} (0.030) & 2.731 (0.023) & 2.815 (0.026) & 2.830 (0.024) \\
\texttt{homes} & \underline{\textbf{0.834}} (0.011) & 0.918 (0.013) & 0.927 (0.015) & 0.912 (0.011) \\
\texttt{imdb\_wiki} & 2.062 (0.040) & 2.037 (0.033) & \textbf{1.954} (0.028) & 2.002 (0.026) \\
\texttt{star} & \textbf{0.240} (0.003) & 0.245 (0.003) & 0.250 (0.004) & 0.258 (0.004) \\
    \bottomrule
  \end{tabular}
\end{table}